\documentclass[12pt]{article}
\usepackage[left=1.in, right=1.in, top=1.in,bottom=1.in]{geometry}

\usepackage{amsmath,amsfonts,amssymb}

\usepackage{soul}

\newcommand{\be}{\begin{equation}}
\newcommand{\ee}{\end{equation}}
\newcommand{\bea}{\begin{eqnarray}}
\newcommand{\eea}{\end{eqnarray}}

\usepackage{ulem}
\usepackage{xcolor}
\usepackage{graphicx}
\usepackage{wrapfig} 
\graphicspath{ {images/} }

\begin{document}

\hfill


\begin{center}
    {\Large\bf CFT Duals for Black Rings and Black Strings} \\
\vspace{2cm}
	{Alexandra Chanson\footnote{E-mail: {\tt a.chanson@usu.edu}}$^{a}$, Victoria Martin\footnote{E-mail: {\tt vlmartin@hi.is}}$^{b}$,}\\
	{ Maria J. Rodriguez\footnote{E-mail: {\tt majo.rodriguez.b@gmail.com}}$^{a,c}$ and Luis Fernando Temoche\footnote{E-mail: {\tt l.f.temoche@usu.edu}}$^{a}$}\\
\vspace{2cm}
{\it a Department of Physics, Utah State University, 4415 Old Main Hill Road, UT 84322, USA}\\
{\it b  University of Iceland, Science Institute, Dunhaga 3, 107 Reykjavik, Iceland}\\
{\it c Instituto de Fisica Teorica UAM/CSIC, Universidad Autonoma de Madrid,13-15 Calle Nicolas Cabrera, 28049 Madrid, Spain}
\end{center}

\vspace{2cm}

\abstract
\vspace{1cm}
%
Holographic dualities between certain gravitational theories in four and five spacetime dimensions and 2D conformal field theories (CFTs) have been proposed based on hidden conformal symmetry exhibited by the radial Klein-Gordon (KG) operator in a so-called near-region limit. In this paper, we examine hidden conformal symmetry of black rings and black strings solutions, thus demonstrating that the presence of hidden conformal symmetry is not linked to the separability of the KG-equation (or the existence of a Killing-Yano tensor).  Further, we will argue that these classes of non-extremal black holes have a dual 2D CFT. 
New revised monodromy techniques are developed to encompass all the cases we consider. 
 
 \newpage

\tableofcontents


\section{Introduction}
The Kerr black hole metric is considered one of the most relevant solutions of General Relativity due to its astrophysical applications, and also celebrated for its mathematical structure and symmetry. The solution describes a rotating asymptotically flat black hole in four spacetime dimensions, and is famously conjectured to be dual to a two dimensional conformal field theory (2D CFT). This conjectured duality holds both in the extremal case (via a near-horizon limit in the extremal Kerr geometry, or NHEK limit) \cite{Guica:2008mu} and the non-extremal case (via a near-horizon limit in the dynamics that reveals  hidden conformal symmetry of the radial Klein-Gordon (KG) equation) \cite{Castro:2010fd}. 

One remarkable feature of the Kerr metric is that it leads to a separable KG-equation \cite{Carter:1968rr}. We will call black hole solutions with a separable KG-equation simply separable black hole metrics. It was later shown that such classes of separable black hole solutions are uniquely determined by requiring the existence of a principal closed conformal Killing-Yano form \cite{Frolov:2006dqt} (and references therein). The geometry possesses a high degree of symmetry encoded in the existence of a tower of Killing vectors and tensors. Unlike Killing vectors, these higher-rank Killing tensors do not generate explicit isometries of the metric, and for this reason they are often called ``hidden'' symmetry generators \cite{Frolov:2017kze}. An interesting question is to determine to what extent these hidden symmetries of separability are related to the hidden \textit{conformal} symmetries found in geometries like Kerr \cite{Castro:2010fd}. 

Many miraculous properties of the Kerr solution survive in higher dimensions. The KG-equations for many higher-dimensional solutions of the Einstein equations are separable, and the geometry possesses a high degree of symmetry encoded in a tower of Killing vectors and tensors \cite{Frolov:2006dqt,Kubiznak:2006kt}. However, this exceptional symmetry does not extend to all higher dimensional black hole solutions of General Relativity. In particular, the KG-equations for generic black ring solutions \cite{Emparan:2006mm} are believed to be non-separable.

In the context of the nonextremal Kerr/CFT correspondence \cite{Castro:2010fd}, hidden conformal symmetry leads to the proposal that D-dimensional black holes are dual to 2D CFTs (in contrast to D$-1$ CFTs) in a certain limit, stemming from the analysis of the separable KG-equations.  As previously mentioned, one pressing question that remains to be clarified is whether the hidden conformal symmetries relevant for the CFT description of nonextremal black holes are bound to the existence of higher degrees of symmetry encoded in Killing tensors, which are ultimately responsible for KG-equation separability. 

We will argue that the CFT interpretation for generic black holes is valid even without full separability of the KG-equation. Moreover, we conjecture that the existence of hidden symmetry structures in the KG-equation will be guaranteed for black hole metrics with two horizons -- an outer event horizon $r=r_+$ and Cauchy inner horizon $r=r_-$. In other words, black hole metrics containing a smooth extremal limit $r_{ext}=r_+=r_-$. This property for black holes seems to be the key for the CFT interpretations rather than separability and existence of Killing-Yano forms. 

Other techniques involving the CFT interpretation of black hole horizons include the so-called monodromy technique \cite{Castro:2013kea,Castro:2013lba,Chanson:2020hly}. In this approach, expressions involving the monodromies are interpreted by making explicit the relationship between the monodromy of the solutions and greybody factors for fully separable KG-equations.  One then wonders: are the solution monodormies, hidden symmetries of KG separability and hidden conformal symmetries from the dynamics related? We find that separability is not required, and therefore not linked to the hidden conformal symmetries in the KG equation. Indeed, the hidden conformal symmetry structure can be obtained from the monodromies, allowing a 2D CFT interpretation even for black holes exhibiting non-separable KG equations. With appropriate modifications to the monodormy technique, we will be able to compute explicit relations that can be interpreted as effective temperatures of a 2D CFT for generic black hole solutions of vacuum Einstein's equations with at least two horizons.

In this article, we pay special attention to two black ring solutions, the dipole black ring \cite{Emparan:2004wy} and the doubly-spinning black ring \cite{pomeransky2006black,Chen:2011jb}. For both solutions, we construct the non-separable KG equation, take a constant angular slice $\theta_o$ (as us done for NHEK), and keep only the pole terms. This approach should be contrasted with the so-called ``near-region limit'' of \cite{Castro:2010fd} and others. In those works, they take a limit $\omega M<<1$, $\omega r<<1$, which has been endowed with a physical interpretation involving soft hair \cite{Haco:2018ske}. However, the same hidden symmetry arguments can be obtained without such a specific limit and simply focusing on the poles of the radial equation. Indeed, we will find that the black ring solutions would require a more complicated limit than that proposed by \cite{Castro:2010fd}.

From the radial pole terms of our KG equations, we can extract the monodromies and build the temperatures $T_L$ and $T_R$ for the purported dual 2D CFT. We then move on to construct conformal coordinates for these spacetimes, to further exhibit the hidden conformal symmetry structure. We show a difficulty arises in constructing an appropriate radial function for the conformal coordinates, reminiscent of what was found in \cite{Keeler:2021tqy} for higher-dimensional spacetimes (D$>$5). We make progress by considering two limits that we describe: a black string limit and a near-horizon limit in the conformal coordinates. In this way we succeed in constructing conformal coordinates for these systems.  

This article is organized as follows. In Section \ref{sec:Methods} we outline our formalism, namely the revised monodromy technique and the construction of conformal coordinates. In Section \ref{sec:Kerr} we demonstrate these techniques on the Kerr black hole before considering our spacetimes of interest: the dipole black ring (Section \ref{sec:Dipole}) and the doubly-spinning black ring (Section \ref{sec:DoubleSpin}). In our Discussion Section \ref{Discussion} we review our results and conclusions. We relegate some metric definitions and properties to Appendix \ref{Appendix:A}.
%


\section{Methods}\label{sec:Methods}
Here we outline some of the techniques that we use to probe hidden conformal symmetry in black ring solutions. While much of this section is review, some of the analysis is, to the best of our knowledge, new. 
\subsection{Revised monodromy technique}

We consider the KG equation for a massless scalar evaluated on a D-dimensional spacetime background 
\be\label{KGini}
\nabla^2 \Psi(x) = {\frac{1}{\sqrt{-g}} } \partial_\mu ( \sqrt{-g}g^{\mu\nu}\partial_\nu)\Psi(x)=0.
\ee
Let us assume that the background has $n+1$ Killing vectors. We choose a basis where the vectors are simply labelled by the coordinates $K=(\partial_t, \partial_{\phi_k})$ with $k=1,\cdots,n$. The presence of Killing vectors allows us to decompose the wave function as
\be\label{fullWV}
\Psi(x)= \exp(-i\omega t + i \sum^{n}_{k=1}   m_{\phi_k}\phi_k)\, \Phi(r,\theta)\,.
\ee
The full separability in $(r,\theta)$ coordinates via $ \Phi(r,\theta)$ does not follow from symmetries realized explicitly in the black hole background and will be the subject of our paper. We will be mostly interested in solutions that have Cauchy horizons, such as rotating black holes and black rings. In the appropriate coordinate system these horizons are characterized by being simple poles of a ``radial'' coordinate. If we denote this coordinate by $r$, the horizons are defined by the zeroes of $g^{rr}=0$. 

The basic observation for the revised monodromy technique starts as follows.  Horizons are regular surfaces, and in particular we expect that the wave equation is well behaved around each pole. Hence, instead of analyzing the behavior of the KG equation and solutions around each pole, we can instead simply keep the kinetic radial part and pole terms for each horizon $\{r_- ,r_+ \}$ to identify the monodromy and corresponding 2D CFT temperatures. This can be regarded as a near zone expansion\footnote{Again, this near zone expansion need not coincide with the near-region limit defined in \cite{Castro:2010fd}.}, with a local hidden $SL(2,R) \times SL(2,R)$ symmetry -- this symmetry does not correspond to an isometry of
the background. 

For example, the near zone limit $r\to  {r_j=\{r_-,r_+\}}$ of the black holes we consider leads to
\be
\partial_r\left[\Delta(r)\partial_r \Phi(r)\right]+ \left[{F_+(\omega, m_{\phi_k})^2\over (r-r_+)}+ {F_-(\omega, m_{\phi_k})^2\over (r-r_-)}\right]\Phi(r) =0,
\ee
where $\Delta(r)= (r-r_+) (r-r_-)f(r)$ for a regular function $f(r)$ at $r_+,r_-$. The solutions to this effective differential equation
with a non-trivial monodromy  $\alpha_j$ around $(r-r_j)$ take the form
\be\label{radial}
\Phi(r) \sim (r-r_j)^{i \, \alpha_j} + (r-r_j)^{-i \alpha_j} ~,\quad \alpha_j = \frac{F_j(\omega, m_{\phi_k})} { \lim_{r\rightarrow r_j} (\Delta(r)/(r-r_j))^{1/2}}.
\ee
The associated monodromy matrix is
\be
M_j=\left(\begin{array}{cc}e^{2\pi \alpha_j} &0\\ 0&e^{-2\pi \alpha_j} \end{array}\right).
\ee
The curiosity here is that we write $m_{\phi_k}$ as a function of $\alpha_j$, such that the coefficients in this relation can be interpreted as effective temperatures of a 2D CFT. More explicitly, take $\phi_k\to \phi_k+2\pi$, and then the wave function changes as \cite{Aggarwal:2019iay}
\be\label{temp2DCFT}
\Psi(\phi_k + 2\pi) = e^{2\pi i m_{\phi_k} } \Psi(\phi_k) = e^{-i4\pi^2 (-T_{L,\phi_k} \omega_L+T_{R,\phi_k} \omega_R) } \Psi(\phi_k)\,,
\ee
where $T_{j,\phi_k}$ is determined by rewriting  \eqref{radial} as  $m_k= m_k(\alpha_j)$. As we explicitly compute below, it seems in general that 
\bea
\omega_{R}=\alpha_++\alpha_-\,,\qquad \omega_{L}=\alpha_+ - \alpha_-,
\eea 
involving the monodromies around the outer (+) and inner (-) horizons. These relations seem to indicate that $T_{R/L,\phi_k}$'s are the relevant temperatures to describe the black hole entropy as a Cardy formula for a 2D CFT. 

\subsection{Building conformal coordinates}
In this subsection we describe a streamlined technique for constructing conformal coordinates, and also express the monodromy parameters $\alpha_{\pm}$ in terms of these. We will use the technique outlined here to attempt to construct coordinates that reproduce the conformal structure of the near zone black ring solutions. 

Following \cite{Castro:2010fd, Perry:2020ndy}, we seek a coordinate transformation of the following form 
\begin{equation}\label{rformconfcoord}
	\begin{split}
		w^+&=f(r)e^{\alpha\phi+\beta t}\\	w^-&=f(r)e^{\gamma\phi+\delta t}\\
		y&=g(r)e^{1/2((\alpha+\gamma)\phi+(\beta+\delta) t)}.
	\end{split}
\end{equation}
Note that we have selected a single angular direction $\phi$. The primary purpose of these conformal coordinates is to construct an $SL(2,R)$ Casimir
\begin{equation}\label{casimir}
	\mathcal{H}^2=\frac{1}{4}\left(y^2\partial_y^2-y\partial_y\right)+y^2\partial_+\partial_-,
\end{equation}
which is proportional to the near zone radial KG operator acting on our probe scalar field $\Phi=R(x)e^{i(k\phi-\omega t)}$. A secondary feature that we would like these coordinates to possess is that near the bifurcation surface ($w^\pm=0$) the metric becomes warped $AdS_3$ to leading order.

It will sometimes be cleaner to reparametrize the radial coordinate $r$ in the following way
\begin{equation}\label{xformconfcood}
	\begin{split}
		w^+&=\sqrt{\frac{x-\frac{1}{2}}{x+\frac{1}{2}}}e^{\alpha\phi+\beta t}\\	w^-&=\sqrt{\frac{x-\frac{1}{2}}{x+\frac{1}{2}}}e^{\gamma\phi+\delta t}\\
		y&=\sqrt{\frac{1}{x+\frac{1}{2}}}e^{1/2((\alpha+\gamma)\phi+(\beta+\delta) t)}.
	\end{split}
\end{equation}
The nice thing about the $x$-coordinate is that this form \eqref{xformconfcood} is the same in both four and five dimensions.  For Kerr, $x=\frac{2r-r_--r_+}{2(r_+-r_-)}$ and for the 5D Myers-Perry black hole $x=\frac{r^2-1/2(r_+^2+r_-^2)}{r_+^2-r_-^2}$. Indeed, we will find that the conformal coordinates for the black ring and black string solutions fit this form as well.  

In terms of these general conformal coordinates \eqref{xformconfcood}, we can write down the Casimir \eqref{casimir} as
\begin{equation}
	\begin{split}
		\mathcal{H}^2R(x)=
		&\left(\partial_x\Delta\partial_x+\frac{(\omega  (\alpha +\gamma )+k (\beta +\delta
			))^2}{4 \left(x-\frac{1}{2}\right) (\beta  \gamma
			-\alpha  \delta )^2}-\frac{(\omega  (\alpha -\gamma
			)+k (\beta -\delta ))^2}{4
			\left(x+\frac{1}{2}\right) (\beta  \gamma -\alpha 
			\delta )^2}\right)R(x),
	\end{split}
\end{equation}
where $\Delta=\left(x-\frac{1}{2}\right)\left(x+\frac{1}{2}\right)$. This is what should be compared to the near zone radial KG operator to determine the parameters $(\alpha,\beta,\gamma,\delta)$. The solutions to the equation $\mathcal{H}^2R(x)=0$ are hypergeometric functions (by construction). This analysis also gives us the monodromy data in terms the the conformal coordinates. The monodromy around the outer horizon is 
\begin{equation}
	R(x)\sim\left(x-\frac{1}{2}\right)^{i\alpha_+}, \qquad \alpha_+=\frac{\omega  (\alpha +\gamma )+k (\beta +\delta
		)}{2  (\beta  \gamma
		-\alpha  \delta )}.
\end{equation} 
The monodromy around the inner horizon is 
\begin{equation}
	R(x)\sim\left(x+\frac{1}{2}\right)^{i\alpha_-}, \qquad \alpha_-=\frac{\omega  (\alpha -\gamma )+k (\beta -\delta
		)}{2  (\beta  \gamma
		-\alpha  \delta )}.
\end{equation} 
These expressions for the monodromies make direct contact with the parameters in the Casimir. We can now identify the CFT temperatures as proposed in \cite{Perry:2020ndy,Perry:2022udk}. The analysis of the periodicities yields
\bea
\alpha= 2\pi T_R,\qquad \gamma=2\pi T_L\,,
\eea
or equivalently
\bea
T_R=\alpha/(2\pi)\,,\qquad T_L=\gamma/(2\pi)\,.
\eea
Via a Cardy formula for a 2D CFT, these will reproduce the expressions for the inner and outer black hole entropies.


\section{Warm-up: Kerr Black Hole}\label{sec:Kerr}

The simplest case where we can exhibit the formalism of Section 2 is the Kerr black hole. For Kerr, with generic mass $M$ and angular momentum $J = M a$, the full radial KG equation is\footnote{For the Kerr metric and more details regarding the resulting KG equation, please see for example \cite{Castro:2010fd}.}
\begin{eqnarray}\label{bbz}
&&\left[\partial_r  \left((r-r_+)(r-r_-)\partial_r\right)
+{({2Mr_+\omega}-{a}m )^2\over (r-r_+)(r_+-r_-)} \right. \cr &&\left. - {({2Mr_-\omega}-{a}m )^2\over (r-r_-)(r_+-r_-)}
+ (r^2+2M(r+2M) )\omega^2 \right] \Phi(r)=K_\ell \,  \Phi(r)~ .\end{eqnarray}
In the near zone expansion we have
\be\label{nearzone}
\left[\partial_r  \left((r-r_+)(r-r_-)\partial_r\right)
+{({2Mr_+ \omega}-{a}m )^2\over (r-r_+)(r_+-r_-)}  - {({2Mr_-\omega}-{a}m )^2\over (r-r_-)(r_+-r_-)} \right] \Phi =0.
\ee
In this case, the near zone expansion \eqref{nearzone} is identical to the near-region limit reported in \cite{Castro:2010fd}. Employing the definition (\ref{radial}), we can now compute the monodromies for the outer horizon $r=r_+$ 
\be\label{n1kerr}
\Phi \sim  (r-r_+)^{\pm \, i\alpha_+} ~,\quad \alpha_+= {({2Mr_+\omega}-{a}m )\over (r_+-r_-)}. 
\ee
Similarly, for the inner horizon $r=r_-$ it gives
\be\label{n2kerr}
\Phi \sim  (r-r_-)^{\pm \, i\alpha_-} ~,\quad \alpha_-= {({2Mr_-\omega}-{a}m )\over (r_+-r_-)}. 
\ee

Solving for $m$ in terms of $\alpha_{\pm}$ we get
\be
-m= {r_+\over a}\alpha_--{r_-\over a}\alpha_+.
\ee
Taking $\phi \to \phi +2\pi$, the wave function (\ref{temp2DCFT}) changes as
\bea
\Psi(\phi+2\pi)= \exp\left(-2\pi i  {r_+\over a}\alpha_- +2\pi i  {r_-\over a}\alpha_+\right)\Psi(\phi).
\eea
Changing the basis to $\omega_{R,L}=\alpha_+\pm \alpha_-$ we get\footnote{This is related to the fact that the identification  $\phi\rightarrow\phi+2\pi$ is generated by the group element $	e^{-i4\pi^2(T_RH_0+T_L\bar{H}_0)}=e^{2\pi\partial_\phi}$.}
\bea\label{aax}
\Psi(\phi+2\pi)= e^{i4\pi^2 \,(T_L\, \omega_L - \,T_R \,\omega_R )}\Psi(\phi),
\eea
where
\be\label{refTemps}
T_L= {r_++r_-\over 4\pi a}~,\quad T_R= {r_+-r_-\over 4\pi a}.
\ee
These are the same temperatures computed in \cite{Castro:2010fd} using the hypergeometric equation.
The CFT microstate degeneracy inferred from the Cardy formula with central charge $c_L=c_R= 12 J$ agrees exactly with the Bekenstein-Hawking area law. Further, the product of the inner and outer entropy, representing a minimal phase-volume, is quantized as a full spin surface area:
\be
S_+ S_- = 4 \pi^2 J^2.
\ee


\section{Dipole Black Ring and Black String}\label{sec:Dipole}


The black ring solutions in this and the upcoming section distinguish themselves from other black objects due to their non-trivial $S^{1}\times S^{2}$ horizon topology. Known as $D=\{10,11\}$ supergravity solutions \cite{Bena:2004de}, they can also be found as systems in minimal  $\mathcal{N}=1, D=5$ supergravity. A classification of those solutions can be found in \cite{Gauntlett:2002nw}. The IR limit of the latter leads to the identification of black rings as exact solutions of General Relativity \cite{Emparan:2001wn} in $D=5$, erecting themselves as clear violations of the uniqueness theorem for black holes \cite{Hawking:1971vc}.
The dipole black ring in particular is a generalization of \cite{Emparan:2001wn}, being an exact solution of Einstein-Maxwell-dilaton theory. We will be considering the case where the dilaton $\phi$ is decoupled, thus giving us pure Einstein-Maxwell theory with winding number $N=3$. More precisely, we will focus on the neutral (non-supersymmetric) dipole spinning black ring 
specified by the three physical parameters  $(M, J_{\psi},q)$: the mass, angular momentum along $S^1$ and dipole magnetic charge, respectively. An in-depth description of the dipole black ring solution is given in Appendix \ref{Appendix:A}. 
The black ring solution \cite{Emparan:2001wn} is parametrized by a scale $R$, the dipole parameter $\mu$ and $\lambda,\nu$ within $ 0<\nu \le \lambda<1$ and $0\le \mu<1$.

Although the $(x,y,\psi)$-coordinates discussed in Appendix \ref{Appendix:A} are useful for a compact expression for the metric, it is instructive to consider $(r,\theta, z)$-coordinates defined by 
\bea
x=\cos\theta , \qquad y=-R/r\,, \qquad \text{ and}  \qquad \psi=z/R\,,
\eea
as well as the constants $\nu=r_0/R$, $\lambda=(r_0 /R) \cosh^2\sigma$ and $\mu= (r_0 /R) \sinh^2\gamma$, where $\gamma$ gives a convenient parametrization of the charge. We label the functions $s^2_X=\sinh^2 X$ and $c^2_X=\cosh^2 X$ from here onwards. This is a convenient set of coordinates to take the straight black string limit $R\rightarrow\infty$.  In these coordinates the outer horizon is at $r=r_0$ and the inner horizon is at $r=0$. The two horizons coincide when $\nu=0$, which defines the extremal limit, and thus $\nu$ can be regarded as a non-extremality parameter. 

The expressions for the entropy were computed in \cite{Castro:2012av} yielding 
\bea\label{DipoleEntropy}
S_+&=&{2\pi^2 r_0^2 R\, c_{\gamma}^3 \,c_{\sigma}\over G_5}{(1+r_0 s^2_{\gamma}/R)^3 (1-(r_0 c^2_{\sigma}/R)^2)^{1/2}\over (1-r_0 /R)^2(1+r_0 /R)}\cr
S_-&=&{2\pi^2 r_0^2 R \, s_{\gamma}^3\, s_{\sigma}\over G_5}{(1+r_0 s^2_{\gamma}/R)^3 (1-(r_0 c^2_{\sigma}/R)^2)^{1/2}\over (1-r_0 /R)^2}\,.
\eea
In addition, we have the angular momentum 
\bea
J_{\psi} = \frac{\pi R^2 r_0 c_{\sigma} s_{\sigma}}{2 G_5} \frac{\left(1+\frac{r_0 s^2_{\gamma}}{R}\right)^{9/2}\left(1+\frac{r_0 c^2_{\sigma}}{R}\right)^{1/2} }{ \left(1-\frac{r_0}{R} \right)^{2}}\,,
\eea
and dipole charge
\be
q^3={2\pi^2 r_0^3 s_{\gamma}c_{\gamma}^3\over G_5}{(1+r_0 s^2_{\gamma}/R)^3\over (1-r_0/R)^3}\left({1-(r_0c^2_{\sigma}/R)\over 1-r_0 s^2_{\gamma}/R}\right)^{3/2}\,.
\ee
Interestingly, as observed in \cite{Castro:2012av}, it is the case that a the product of the inner and outer entropies is independent of the mass of the black hole
and therefore depends solely on the quantized charges
\be
S_+ S_- = 4 \pi^2 \, J_{\psi} \, q^3.
\ee

We now use the revised monodromy technique to analyze the KG-equation for a massless scalar field on the dipole black ring geometry in the $(r,\theta,z)$-coordinates. This allows us to compute temperatures of the proposed dual 2D CFT.

\subsection{Monodromy analysis}
In eigenmodes the scalar field is of the form
\bea
\Psi(t,r,\theta,\phi,z)=e^{-i t \omega+i m \phi+i n z} \,\left(1+\frac{r}{R}\cos\theta\right) \, \Phi(r,\theta).
\eea
Notice that both $m$ and $n$ are not integers because $(\phi,z)$ do not have periodicity $2\pi$. Below we will account for this. The classical wave equation for $\Phi(r,\theta)$ becomes
\bea\label{KGring}
&&\partial_r \left( r\,(r-r_0)\left(1-\frac{r^2}{R^2}\right)\partial_r\, \Phi \right)+\frac{1}{\sin\theta}\,\partial_\theta \left(\left(1+\frac{r_0}{R}\cos\theta\right)\sin\theta \,\partial_\theta\, \Phi \right)\\
&&+\frac{(r+r_0 \,s_\gamma^2 )^3 }{r\, (r_0-r)\left(1-\frac{r^2}{R^2}\right)(r-r_0 c_\sigma^2)}\left(\omega \, r_0 c_\sigma s_\sigma \left(1-\frac{r}{R}\right)\sqrt{\frac{1+\frac{r_0}{R}c_\sigma^2}{1-\frac{r_0}{R}c_\sigma^2}}-n (r-r_0 c_\sigma^2)\right)^2  \Phi \nonumber\\
&&+\omega^2\frac{(1+\frac{r_0}{R}c_\sigma^2\cos\theta)^2(r+r_0 \,s_\gamma^2)^3}{(1+\frac{r}{R}\cos\theta)^2(r-r_0c_\sigma^2)} \, \Phi -m^2\frac{(1+\frac{r_0}{R}c_\sigma^2\cos\theta)(1-\frac{r_0}{R}s_\gamma^2\cos\theta)^3}{(1+\frac{r_0}{R}\cos\theta)\sin^2\theta} \,\Phi\nonumber \\
&&+\left( \frac{r_0}{R}\,\cos\theta-\frac{r}{R}(2 r-r_0)\right )\,\Phi =0\nonumber
\eea
%
 If we only keep the kinetic radial part and the single poles for each horizon, the wave function becomes
\bea\label{SimpleDipoleKG}
&&\partial_r \left( {r\,(r-r_0)(1-\frac{r^2}{R^2})}\,\partial_r\, \Phi\right)+ {r_0^3\over r-r_0}c_\gamma^6{1-{r_0\over R}\over 1+{r_0\over R}} \left({s_\sigma \over 1-{r_0\over R}}n+{c_\sigma} K \omega\right)^2  \Phi\\
&&-{r_0^3\over r}s_\gamma^6\left(c_\sigma n+ s_\sigma K \omega\right)^2  \Phi   =0.\nonumber
\eea

The calculation of the monodromies employing the prescription (\ref{radial}) is straightforward. Around  the outer event horizon $r \to r_0$:
\be
\Phi \sim (r-r_0)^{\pm  i\alpha_+} ~,\quad \alpha_+={r_0}{c_\gamma^3\over 1+{r_0\over R}} \left({s_\sigma \over 1-{r_0\over R}}n+ {c_\sigma} K\omega\right).
\ee
We can also compute the monodromy around the inner horizon $r\to 0$:
%
\be
\Phi \sim r^{\pm i\alpha_-} ~,\quad \alpha_-={r_0}s_\gamma^3\left(c_\sigma n+ s_\sigma K \omega\right).
\ee

Here we have an expression for $\alpha_\pm$ as a function of $\omega$ and $n$. Solving for $n$ gives
\be
n= {1-{r_0\over R}\over 1-c_\sigma^2{r_0\over R}}{1\over 2 r_0 s_\gamma^3 c_\gamma^3}\left[ \left(c_\sigma c_\gamma^3-s_\sigma s_\gamma^3(1+{r_0\over R})\right)\, \omega_L +\left(c_\sigma c_\gamma^3+s_\sigma s_\gamma^3(1+{r_0\over R})\right)\, \omega_R\right]
\ee
with $\omega_{R,L}=\alpha_+\pm \alpha_-$. The wave function transforms as 
\bea
\Psi(z+2\pi  \Delta z)= e^{2\pi \Delta z \, i n } \Psi(z) = e^{-i4\pi^2\left(-T_{L} \omega_L +T_{R} \omega_R \right)}\Psi(z)
\eea
with conformal temperatures
\bea\label{DipoleTemps}
T_R&=& {\Delta z\over 2\pi} {1-{r_0\over R}\over 1-c_\sigma^2{r_0\over R}}{1\over 2 r_0 s_\gamma^3 c_\gamma^3}\left(c_\sigma c_\gamma^3-s_\sigma s_\gamma^3(1+{r_0\over R})\right)\cr
T_L&=& {\Delta z\over 2\pi} {1-{r_0\over R}\over 1-c_\sigma^2{r_0\over R}}{1\over 2 r_0 s_\gamma^3 c_\gamma^3}\left(c_\sigma c_\gamma^3+s_\sigma s_\gamma^3(1+{r_0\over R})\right)
\eea
and
\bea
\Delta z= 2\pi \frac{(1+r_0 s_{\gamma}^2/R)^{3/2}(1- r_0 c_{\sigma}^2/R)^{1/2}}{(1-r_0/R)}\,.
\eea
Here $\Delta z$ is the periodicity of the azimuthal direction in the black ring solution. Fixing this condition represents a balance between forces in the ring that can be achieved when there are no conical
singularities. See e.g. \cite{Emparan:2004wy} for more details. Our results here for the dipole black ring apply for both situations, with or without balance. For our computations, in order to link the results between the black ring and strings, it will be useful to not fix the bound. Note that to be able to take the black string limit $R\rightarrow \infty$ one needs to to keep the quantity $\Delta z$  unfixed.
 
%
A straightforward calculation using these results shows that the black ring satisfies a Cardy entropy formula \cite{Cardy:1986ie}:
\bea
S_{\pm}&=&\frac{c\pi^2}{3}( \,T_{L} \pm \, T_{R})\,
\eea
with central charge
\bea\label{centralNonSup}
c=6\,q^3\,
\eea 
and the entropies reported in \eqref{DipoleEntropy}. In analogy to the supersymmetric cases \cite{Emparan:2006mm}, the entropy is independent of the mass and a power of the dipole charge. The central charge agrees with that of \cite{Emparan:2004wy}. In these cases the Bekenstein-Hawking entropy of a large extremal ring can be reproduced through a microscopic calculation.

It is worth noticing that the extremal black ring configuration $\nu\rightarrow 0$ has $T_R=0$. The extremal system $\nu=0$ is regarded as the ground state of the ring with finite radius. The matching to the results previously obtained from the thermodynamical characterization of the dual CFT is remarkable. In the supersymmetric case, the black ring exhibits a set of three dipole charges $q_i$, which when quantized become $n_i$:
\begin{equation}
	n_i=\Big(\frac{\pi}{4G_{5}}\Big)^{1/3}q_i.
\end{equation}

  For both dipole black ring and its supergravity cousin, extremality is reached when $\nu \rightarrow 0$. For the latter case, the extremal configuration can be regarded as the ground state of a ring with large but finite radius \cite{Emparan:2004wy}.  
  Then in the extremal, large string limit, the central charge becomes
\begin{equation}
	\label{eqn:centralstringsugra}
	c=6{n}_{1}{n}_{2}{n}_{3}{n}_{p}.
\end{equation}
Further, for these configurations $n_p$ corresponds to the chiral momentum excitations. Determined through a Komar integral, its value  is:
\begin{equation}
	\label{eqn:chiralring}
	n_p=J
\end{equation}
where $J$ is the ADM value of the angular momentum of the ring and $n_i$ can be identified with the number of each type of M5 branes forming the ring, 
\begin{equation}
\label{eqn:chiralstring}
n_i=\left(\frac{2\pi}{G}\right)^{1/3}Q_i\,.
\end{equation}


\subsection{Boosted charged black string}\label{subsec:BoostedChargedString}

The black string solution appears as a higher dimensional solution of Einstein-Maxwell theory. These black objects attracted great interest regarding their event horizon stability. The well-known Gregory-Laflamme instability \cite{Gregory:1993vy} was described first in the context of black strings. Since then, further studies of the instability and the fate of black strings evolving into regular higher dimensional black hole configurations have populated the literature \cite{Wiseman:2002ti,Horowitz:2001cz}.

We introduce the metric and gauge field for the charged boosted black string 
\begin{eqnarray}
    \label{eqn:CBBSmetric}
    &&ds^2=-\frac{\hat{f}}{h}\Big[dt-\frac{{r_0}{c_{\sigma}}{s_{\sigma}}}{r\hat{f}}dz\Big]^{2}+\frac{f}{\hat{f}h}dz^{2}+h^{2}\Big[\frac{dr^{2}}{f}+r^{2}d{\Omega}^{2}_{2}\Big]\\
    &&A_{\phi}=\sqrt{3}{r_0}s_{\gamma}c_{\gamma}(1+\cos({\theta})),
\end{eqnarray}
where ${d\Omega_2}^{2}={d\theta}^{2}+{\sin{\theta}}^{2}{d\phi}^{2}$ and
\begin{eqnarray}
    &&f=1-\frac{r_0}{r}\\
    &&{\hat{f}}=1-\frac{{r_0}{c_{\sigma}}^{2}}{r}\\
    &&h=1+\frac{{r_0}{s_{\gamma}}^{2}}{r}.
\end{eqnarray}
In the metric \eqref{eqn:CBBSmetric} the coordinates that we will use to construct our conformal coordinates will be $(t,r,z)$. The entropy of the outer and inner horizons is given by \cite{Compere:2010fm}
\begin{equation}\label{DipoleStringEntropies}
	S_{+}=\frac{2{\pi}^{2}{r}_{0}^{2}R{c}_{\gamma}^{3}c_{\sigma}}{G_5}, \qquad S_{-}=\frac{2{\pi}^{2}{r}_{0}^{2}R{s}_{\gamma}^{3}s_{\sigma}}{G_5}.
\end{equation} 

By introducing the coordinate transformation $x=\frac{r}{r_0}-\frac{1}{2}$ on the metric, it is possible to identify the outer/inner horizon at $x=\pm \frac{1}{2}$. Then, we calculate the KG equation of the system using the ansatz
\begin{equation}
    \label{eqn:ansatzChargedString}
    \Psi(t,x,\theta,\phi,z)=e^{-it\omega+im\phi+inz}R(x)\Omega(\theta)
\end{equation}
and we verify that the equation is separable. The radial part is given by: 
\begin{eqnarray}
    \label{eqn:KGchargedstring}
    &&\Big[{\partial}_{x}[\Delta{\partial}_{x}]\\\nonumber
    &&+\frac{r_{0}^{2}(c_{\gamma}^{2}+s_{\gamma}^{2}+2x)^{3}[{\omega}^{2}(c_{\sigma}^{2}+s_{\sigma}^{2}+2x)+n^{2}(c_{\sigma}^{2}+s_{\sigma}^{2}-2x)+4{\omega}nc_{\sigma}s_{\sigma}]}{16(x-1/2)(x+1/2)}\Big]R(x)\nonumber\\
    &&=K_{l}R(x)\nonumber
\end{eqnarray}
where $\Delta=(x-\frac{1}{2})(x+\frac{1}{2})$. The angular component is
\begin{eqnarray}
    \frac{1}{\sin{\theta}}{\partial}_{\theta}\Big[\sin\theta{\partial}_{\theta}\Omega\Big]-\frac{m^2}{{\sin}^{2}\theta}\Omega=-K_{l}\Omega.
\end{eqnarray}
We will focus on the radial equation and isolate the pole terms in (\ref{eqn:KGchargedstring}). The wave equation becomes:
\begin{equation}
    \Big[{\partial}_{x}[\Delta{\partial}_{x}]+{r_0}^{2}\Big(\frac{{c_{\gamma}}^{6}}{x-\frac{1}{2}}(ns_{\sigma}+\omega c_{\sigma})^{2}
    -\frac{{s_{\gamma}}^{6}}{x+\frac{1}{2}}(nc_{\sigma}+\omega s_{\sigma})^{2}\Big)\Big]R(x)=0.
\end{equation}
This agrees with the $R\rightarrow\infty$ limit of \eqref{SimpleDipoleKG}.
\\
 
Around the inner horizon $r\rightarrow 0$, the monodromy is given by:
\begin{equation}
	\Phi \sim r^{\pm i\alpha_-} ~,\quad \alpha_-=r_{0}{s}_{\gamma}^{3}(n {c}_{\sigma}+\omega {s}_{\sigma})
\end{equation}
and with respect to the outer horizon $r\rightarrow r_0$:
\begin{equation}
	\Phi \sim r^{\pm i\alpha_+} ~,\quad \alpha_+=r_{0}{c}_{\gamma}^{3}(n {s}_{\sigma}+\omega {c}_{\sigma}).
\end{equation}
Solving for $n$ gives
\begin{equation}\label{nequation}
	n=\frac{1}{2{r}_{0}{c}_{\gamma}^{3}{s}_{\gamma}^{3}}\Big[-({c}_{\sigma}{c}_{\gamma}^{3}+{s}_{\sigma}{s}_{\gamma}^{3}){\omega}_{L}+({c}_{\sigma}{c}_{\gamma}^{3}-{s}_{\sigma}{s}_{\gamma}^{3}){\omega}_{R}\Big]
\end{equation}
with $\omega_{R,L}=\alpha_+\pm \alpha_-$. 
The wave function transforms as
\begin{eqnarray}
	\Psi(z+2\pi  \Delta z)= e^{2\pi \Delta z \, i n } \Psi(z) = e^{\left(i4\pi^2T_{L} \omega_L -i4\pi^2T_{R} \omega_R \right)}\Psi(z).
\end{eqnarray}
We can read off the left and right temperatures 
\be
T_L=\frac{\Delta z}{2\pi}\frac{({c}_{\sigma}{c}_{\gamma}^{3}+{s}_{\sigma}{s}_{\gamma}^{3})}{2{r}_{0}{c}_{\gamma}^{3}{s}_{\gamma}^{3}}, \qquad T_R=\frac{\Delta z}{2\pi}\frac{({c}_{\sigma}{c}_{\gamma}^{3}-{s}_{\sigma}{s}_{\gamma}^{3})}{2{r}_{0}{c}_{\gamma}^{3}{s}_{\gamma}^{3}}.
\ee
Again, assuming a Cardy formula holds, we  recover the entropies \eqref{DipoleStringEntropies} via completely different means. The central charge is in this case $c= 6 q^3$, with
\be
q^3={2\pi^2 r_0^3 s_{\gamma}c_{\gamma}^3\over G_5}.
\ee

\subsection{Conformal coordinates}
We attempt to find conformal coordinates of the form \eqref{rformconfcoord} that reproduce the near zone, constant theta slice Laplacian for the dipole black ring. Our first objective is to determine suitable radial dependence of the functions $f(r)$ and $g(r)$. The procedure that we follow is outlined in Section 3.2 of \cite{Keeler:2021tqy}. Defining the function $h(r)\equiv f(r)/g(r)$, the authors of \cite{Keeler:2021tqy} found that the Casimir operator in Boyer-Lindquist coordinates takes the form 
\be
\begin{split}
	\mathcal{H}^2 = &\frac{h^2+1}{4(h')^2}\partial_r^2 + \left(\frac{1+h^2}{4hh'}\left[\frac{(h')^2-h''h}{(h')^2}\right]+\frac{h}{2h'}\right)\partial_r+F_1(r)\partial_t^2
	+F_2(r)\partial_t\partial_\phi+
	F_3(r)\partial_\phi^2.
\end{split}
\ee
In comparing this to the near zone KG operator \eqref{SimpleDipoleKG}, we can determine the radial dependence of the conformal coordinates $h(r)$ by solving the following differential equation
\begin{equation}
	\frac{ \frac{ h^2+1}{4 h'h}\frac{d}{dr}\left(\frac{h}{h'}\right)+\frac{h}{2
			h'}}{\frac{h^2+1}{4 h'^2}}=\frac{\frac{d\Delta}{dr}}{\Delta},
\end{equation} 
where for the dipole black  $\Delta=r
(r-r_0)\left(1-\frac{r^2}{R^2}\right)$. This has the following solution
\begin{equation}\label{GenHDipole}
	h(r)=	\frac{ \exp \left(c_2-c_1 \left(\frac{\log
			(r-r_0)}{r_0^3-R^2 r_0}+\frac{\log (r)}{R^2
			r_0}+\frac{\log (r-R)}{2 R^2 (R-r_0)}-\frac{\log
			(r+R)}{2 R^2 (R+r_0)}\right)\right)}{\sqrt{1-\exp
			\left(2 \left(c_2-c_1 \left(\frac{\log
				(r-r_0)}{r_0^3-R^2r_0}+\frac{\log (r)}{R^2
				r_0}+\frac{\log (r-R)}{2 R^2 (R-r_0)}-\frac{\log
				(r+R)}{2 R^2 (R+r_0)}\right)\right)\right)}},
\end{equation}
where $c_1$ and $c_2$ are constants. We see here that we are not able to choose a good $c_1$ that eliminates all of the branch cut behavior for us. This is the same problem that occurs in higher dimensions, as pointed out in \cite{Keeler:2021tqy}: a solution $h(r)$ that is free of branch cuts (besides the overall square root) can only exist if the highest power of $\Delta\sim r^2$ (or in the special case where one can write $\Delta\propto(r^2-r_+^2)(r^2-r_-^2)$).

There are two ways to make progress here. The first is to focus only on the outer horizon and take a near horizon limit in the conformal coordinates themselves. This is a reasonable thing to do, as it essentially is taking a near horizon limit in the dynamics, which is close in spirit to the standard hidden conformal symmetry analyses \cite{Castro:2010fd,Haco:2018ske}. To do this, we choose the constants in equation \eqref{GenHDipole} to be $c_2=0$ and $c_1=r_0(R^2-r_0^2)/2$, in analogy to what was suggested in \cite{Keeler:2021tqy}. Then we expand the result around $r=r_0$ to obtain
\begin{equation}\label{dipoleH}
h(r)=	f(r_0,R)\sqrt{\frac{r-r_0}{r_0}},
\end{equation}
where $f(r_0,R)$ is an uninteresting function that is independent of $r$. You will notice that this is reminiscent of the Kerr answer $h(r)=\sqrt{\frac{r-r_+}{r_+-r_-}}$, with $r_+=r_0$ and $r_-=0$.

The second way to make progress is to consider the black string limit $R\rightarrow\infty$. In that case, we would like to solve the differential equation
\begin{equation}
	\frac{ \frac{ h^2+1}{4 h'h}\frac{d}{dr}\left(\frac{h}{h'}\right)+\frac{h}{2
			h'}}{\frac{h^2+1}{4 h'^2}}=\frac{2r-r_0}{r
		(r-r_0)}.
\end{equation} 
This has the solution 
\begin{equation}
	h(r)= \frac{ e^{c_2} r^{\frac{c_1}{r_0}}}{\sqrt{(r-r_0)^{\frac{2
					c_1}{r_0}}-e^{2 c_2}
			r^{\frac{2 c_1}{r_0}}}}.
\end{equation}
We have freedom to choose $c_1$ and $c_2$. One clean choice is to take $c_1=r_0/2$ and $c_2=0$. Then our radial functions take the particularly simple form 
\begin{equation}\label{radialfunctions}
	h(r)=	\sqrt{-\frac{r}{r_0}}, \qquad g(r)= \sqrt{-\frac{r_0}{r-r_0}},\qquad f(r)=gh= \sqrt{\frac{r}{r-r_0}}.
\end{equation} 
The fact that two of these radial functions are imaginary does not cause any problems. The important thing is that the Casimir 
\begin{equation}
	\mathcal{H}^2=\frac{1}{4}\left(y^2\partial_y^2-y\partial_y\right)+y^2\partial_+\partial_-
\end{equation}
is still real. The radial functions \eqref{radialfunctions} admit the same $x$-coordinate structure as presented in \eqref{xformconfcood}. For the dipole black ring, we see that $x=\frac{r}{r_0}-\frac{1}{2}$. 

At last we are ready to attempt to find the angular conformal coordinate variables $(\alpha,\beta,\gamma,\delta)$. To do this, we would like to compare the Casimir 
\begin{equation}\label{CasimirConfCoord}
	\begin{split}
		\mathcal{H}^2R(x)=
		&\left(\partial_x\Delta\partial_x+\frac{(\omega  (\alpha +\gamma )+n (\beta +\delta
			))^2}{4 \left(x-\frac{1}{2}\right) (\beta  \gamma
			-\alpha  \delta )^2}-\frac{(\omega  (\alpha -\gamma
			)+n (\beta -\delta ))^2}{4
			\left(x+\frac{1}{2}\right) (\beta  \gamma -\alpha 
			\delta )^2}\right)R(x)
	\end{split}
\end{equation}
and the $R\rightarrow\infty$ limit of the dipole black ring solution \eqref{SimpleDipoleKG}, written in the $x$-coordinate:
\bea\label{KGDipoleString}
&&\partial_x \left( {\Delta}\,\partial_x\,\Phi \right)+ r_0^2\left({c_\gamma^6\over x-\frac{1}{2}} \left({s_\sigma}n+{c_\sigma}  \omega\right)^2  -{s_\gamma^6\over x+\frac{1}{2}}\left(c_\sigma n+ s_\sigma  \omega\right)^2   \right)\Phi  =0,
\eea
where $\Delta=\left(x-\frac{1}{2}\right)\left(x+\frac{1}{2}\right)$. We can compare $\omega$ and $n$ terms in equations \eqref{CasimirConfCoord} and \eqref{KGDipoleString} to find the conformal coordinate parameters $(\alpha, \beta,\gamma,\delta)$, as in  \eqref{xformconfcood}:
\begin{equation}\label{confcoordagain}
	\begin{split}
		w^+&=\sqrt{\frac{x-\frac{1}{2}}{x+\frac{1}{2}}}e^{\alpha z+\beta t}\\	w^-&=\sqrt{\frac{x-\frac{1}{2}}{x+\frac{1}{2}}}e^{\gamma z+\delta t}\\
		y&=\sqrt{\frac{1}{x+\frac{1}{2}}}e^{1/2((\alpha+\gamma)z+(\beta+\delta) t)}.
	\end{split}
\end{equation}
Due to the squared terms in \eqref{CasimirConfCoord} and \eqref{KGDipoleString}, there are 16 possible combinations of $(\alpha, \beta,\gamma,\delta)$ that work. We choose a branch based on two criteria. First, we would like $\alpha=2\pi T_R$ and $\gamma=2\pi T_L$, where $T_R$ and $T_L$ are given by the $R\rightarrow\infty$ limit of \eqref{DipoleTemps}. Second, when we plug the conformal coordinates \eqref{confcoordagain} into the black string metric \eqref{eqn:CBBSmetric}, we would like to see a warped AdS$_3$ factor near the bifurcation surface $w^\pm=0$. The result is 
\begin{equation}
\begin{split}
		\alpha=\frac{(c_{\sigma}\,c_{\gamma}^3-s_{\sigma}\,s_{\gamma}^3)}{2r_0 \,s_{\gamma}^3 c_{\gamma}^3}\,\qquad \beta= \frac{(- c_{\sigma}\,s_{\gamma}^3+ s_{\sigma}\,c_{\gamma}^3)}{2r_0 \,s_{\gamma}^3 c_{\gamma}^3}\\
	\gamma=\frac{(s_{\sigma}\,s_{\gamma}^3 + c_{\sigma}\,c_{\gamma}^3)}{2r_0 \,s_{\gamma}^3 c_{\gamma}^3}\qquad \delta=\frac{( c_{\sigma}\,s_{\gamma}^3+ s_{\sigma}\,c_{\gamma}^3)}{2r_0 \,s_{\gamma}^3 c_{\gamma}^3}\ .
\end{split}
\end{equation} 
As desired, the metric near the bifurcation surface has a warped AdS$_3$ factor:
\begin{equation}\label{NearBifMet}
	ds^2=4r_0^2c_\gamma^4\left(\frac{dw^+dw^-+\frac{s_\gamma^6}{c_\gamma^6}dy^2}{y^2}+...\right).
\end{equation}
The correction terms in \eqref{NearBifMet} are at least second order in $w^\pm$.

\section{Doubly-spinning Black Ring and  Kerr Black String}\label{sec:DoubleSpin}

In this section we study the hidden conformal symmetries of the doubly-spinning black ring. The metric is a vacuum solution to the five-dimensional Einstein equations with event horizon topology  $S^1 \times S^2$.  The regular solution was constructed by Pomeransky and Senkov \cite{pomeransky2006black} and is parametrized by mass and two angular momenta $(M, J_{\phi},J_{\psi})$ respectively. It is a single black ring configuration balanced by angular momentum $J_{\psi}$ in the plane of the ring, but with angular momentum $J_{\phi}$ also in the orthogonal plane, corresponding to rotation of the $S^2$ sphere (see \cite{Elvang:2007hs} for a detailed analysis of the physical properties). A more general version of the doubly-spinning black ring solution, corresponding to an {\it unbalanced} ring with conical singularities was later found in \cite{Chen:2011jb}. It is in fact this more general {\it unbalanced} black ring solution that contains the Kerr black string and 5-dimensional Myers-Perry black hole as a “collapse” limit of the balanced ring solution. To keep our analysis as general as possible we obtain our results for the solutions \cite{Chen:2011jb}.

In order to set up the notation, we give some details of the doubly spinning black ring in Appendix \ref{Appendix:DSBR}. The black ring solution has been given in different related $(x,y)$-forms. We can rewrite this same foliation of space in a manner that is particularly appropriate in the region near the ring. The $(r,\theta)$-coordinates employed are a transformation from the $(x,y)$-coordinates where 
\bea
x=\cos\theta , \qquad y=-R/r\,, \qquad \text{ and}  \qquad \psi=-z/R\,,
\eea
and has four independent physical parameters $\mu,\nu,\lambda, R$, of which the first three are
dimensionless and the last sets the scale of the solution which are required to satisfy
\bea\label{constrain}
0\le \nu \le \mu\le \lambda < 1 \,.
\eea
The single spinning black ring \cite{Emparan:2006mm} with rotation in only one plane is found in the limit $\nu \rightarrow 0$. The radial and angular coordinates take the ranges $0\le r<R$ and $0\le  \theta< \pi$.\footnote{One can also redefine the constants as $\nu=(M-\sqrt{M^2-a^2})/R= r_-/R$, $\mu=(M+\sqrt{M^2-a^2})/R= r_+/R$ and $\lambda=r_+c_{\sigma}/R$. Taking the ring radius $R$ much larger than the ring thickness $\mu R$ gives the boosted Kerr-black string metric.Note that $\sqrt{2} R=\varkappa$ with respect to the definitions of \cite{Chen:2011jb}.}. The KG equation of the doubly-spinning black ring solution has a hidden phase space symmetry that we argue can be linked to a 2D CFT.

\subsection{Monodromy analysis}
We now turn to a general analysis of the KG-equation \eqref{KGini} for a massless scalar with the ansatz
\be
\tilde\Psi(t,r,\theta,\phi,z)=e^{-i t \omega+i m \phi+i n z} \,\left(1+\frac{r}{R}\cos\theta\right)\,\tilde\Phi(r,\theta)
\ee
in the background of the doubly spinning black ring \cite{pomeransky2006black,Chen:2011jb}
\bea\label{KGPSring}
&&\partial_r \left(\left(1-\frac{r^2}{R^2}\right)(r-\mu\,R)(r-\nu\,R)\,\partial_r\, \tilde\Phi \right)\\
&&+\frac{1}{\sin\theta}\,\partial_\theta \left((1+\mu\cos\theta)(1+\nu\cos\theta)\sin\theta\,\partial_\theta\, \tilde\Phi \right)\nonumber\\
&&+\left(\frac{C_{\mu\nu\lambda} (1-\mu\nu)^2 \left(1+\frac{r}{R}\cos\theta\right)^2 (r/R)^2 \,\tilde{K}(r,\theta) }{\left(1-\frac{r^2}{R^2}\right)(r-\mu\,R)(r-\nu\,R) (1+\mu\cos\theta)(1+\nu\cos\theta)\sin^2\theta }+\tilde f_r+\tilde f_{\theta}\right)\tilde\Phi =0\nonumber
\eea
where 
\bea
\tilde{K}(r,\theta)&=&n^2 R^2 F_{r\theta}-m^2 F_{\theta r}+2 \,m \,\omega (F_{\theta r}\,\omega_{\phi}-J_{r\theta}\,\omega_{\psi})+2 \,n\,R \,\omega (F_{r\theta }\,\omega_{\psi}+J_{r\theta}\,\omega_{\phi})\nonumber\\
&&- 2\,n\, m\,R \,J_{r\theta}+\omega^2\left(\frac{H_{\theta r}^2}{H_{r\theta}}\,\tilde{F}^2+F_{r\theta}\,\omega_{\psi}^2-F_{\theta r}\,\omega_{\phi}^2+2\,\omega_{\phi}\,\omega_{\psi}\,J_{r\theta}\right)
\eea
\bea
\tilde{F}\equiv\left(\frac{F_{\theta r} F_{r\theta } +J_{r\theta}^2}{H_{\theta r} H_{r \theta} }\right)^{1/2}={\left(r^2-R^2\right)^{1/2}(r-\mu\,R)^{1/2}(r-\nu\,R)^{1/2}G(\cos\theta)^{1/2}\over(1-\nu\mu) (1+{r\over R}\cos\theta)^2}
\eea
\be
 C_{\mu\nu\lambda}=\frac{(1-\mu)^2(1-\nu)}{(1-\mu\nu)(1-\lambda)\Phi\Psi}
\ee
\be
F_{r\theta}\equiv F(-R/r,\cos\theta)\,,\qquad F_{\theta r}\equiv F(\cos\theta,-R/r)
\ee
\be
H_{r\theta}\equiv H(-R/r,\cos\theta)\,,\qquad H_{\theta r}\equiv H(\cos\theta,-R/r)
\ee
\be 
J_{r\theta}\equiv J(-R/r,\cos\theta)
\ee
\bea
\tilde f_r=\frac{(\mu+\nu)\,r}{R}-\frac{2 \,r^2}{R^2}\qquad \tilde f_{\theta}=-\cos\theta\,(\mu+\nu+2\,\mu\,\nu\cos\theta)\,,
\eea
and the definitions of $F,G,H,J, \Phi,\Psi,\Sigma,\omega_{\psi},\omega_{\phi}$ can be found in Appendix \ref{Appendix:DSBR}.
Recall that the doubly spinning black ring satisfies the constraints \eqref{constrain} ensuring that the quantities $\Phi$, $\Psi$ and $\Sigma$ are positive. Under these conditions, the solution has a regular event horizon located at $r=\mu\, R$. In addition, there is an inner Cauchy horizon at $r=\nu\, R$.  The two horizons coincide when $\nu=0$, which defines the extremal limit, hence $\nu$ can be regarded as a non-extremality parameter. In general, a ring-shaped ergosurface is present at $r=\lambda R$.

Focusing on a constant $\theta$-slice of the region near the ring $r\rightarrow \{ \mu\,R, \nu \,R \}$ while maintaining all the terms in the  KG-equation that have a singular behavior at the location of the horizon, we find that the equation (\ref{KGPSring}) takes the form
\bea\label{DSRadial}
&&\partial_r \left(\left(1-\frac{r^2}{R^2}\right)(r-\mu\,R)(r-\nu\,R)\,\partial_r\, \tilde\Phi \right)\\
&&-\frac{ C_{\mu\nu\lambda}\, (1-\mu\nu)}{(r-\mu \,R )(\mu-\nu)^2 }\left(\sqrt{\frac{\nu  \Sigma  \Psi }{ \Phi }}\,m-\frac{  (\mu +\nu ) \Sigma\,R }{(1+\mu)} \sqrt{\frac{\lambda  (1+\lambda )}{(1-\mu  \nu )(1-\lambda )}}\,\omega\right)^2\tilde\Phi \nonumber\\
&&-\frac{C_{\mu\nu\lambda}\,(\mu+\nu)^2 (1-\mu\nu)(1-\lambda^2) \nu^2}{(r-\nu \,R )(1-\nu^2)(\mu-\nu)^2 }\times\nonumber\\
&&\left(\frac{\mu\,(1-\lambda \mu)+\nu^2(\lambda-\mu)}{(\mu+\nu)}\sqrt{\frac{1}{ \nu\Phi }}\,m- (1-\mu)R\sqrt{\frac{\lambda (1+\lambda)  \Sigma}{(1-\lambda)(1-\mu  \nu )\Psi }}\,\omega\right)^2\tilde\Phi=0\nonumber
\eea
for the $n=0$ sector in with analogy to the Myers-Perry black hole solution in \cite{Castro:2013lba,Castro:2013kea}. We find that taking $m=0$ is not possible. This is in contrast with the Myers-Perry black hole solution, that is $\phi,\psi$ symmetric.

The regular singular points of the equation (\ref{DSRadial}) correspond to the outer horizon $r=\mu\, R$, inner horizon $r=\nu\, R$ and $r=\pm R$. Now we can proceed with the monodromy technique in the usual way. The solutions to this effective differential equation have a non-trivial monodromy  $\alpha_{\pm}$ around each horizon. 
Near the outer horizon $r\rightarrow \mu\,R$ we have
\be
\tilde\Phi\sim(r-\mu \,R)^{\alpha_+}
\ee
\bea
 \alpha_+&=&\sqrt{\frac{ C_{\mu\nu\lambda}\, (1-\mu\nu)}{(\mu-\nu)^2 }}\left(\frac{  (\mu +\nu ) \Sigma\,R }{(1+\mu)} \sqrt{\frac{\lambda  (1+\lambda )}{(1-\mu  \nu )(1-\lambda )}}\,\omega-\sqrt{\frac{\nu  \Sigma  \Psi }{ \Phi }}\,m\right).\nonumber
\eea
Around the inner Cauchy horizon $r\rightarrow \nu\,R$ we find
\be
\tilde\Phi\sim(r-\nu \,R)^{\alpha_-}\,,
\ee
where
\bea
 \alpha_-&=&\sqrt{\frac{C_{\mu\nu\lambda}\,(\mu+\nu)^2 (1-\mu\nu)(1-\lambda^2) \nu^2}{(1-\nu^2)(\mu-\nu)^2 }}\times\\
&& \left((1-\mu)R\sqrt{\frac{\lambda (1+\lambda)  \Sigma}{(1-\lambda)(1-\mu  \nu )\Psi }}\,\omega-\frac{\mu\,(1-\lambda \mu)+\nu^2(\lambda-\mu)}{(\mu+\nu)}\sqrt{\frac{1}{ \nu\Phi }}\,m\right).\nonumber
\eea

To analyze the temperatures of the proposed dual CFT we are obliged to consider each spin sector separately in analogy to \cite{Krishnan:2010pv}. We begin writing the above monodromy parameters as
\begin{equation}
	\alpha_+=A\omega+Bm, \qquad \alpha_-=C\omega+Dm.
\end{equation}
Solving for $\omega$ and $m$, we find
\begin{equation}\label{meqDBR}
	m=\frac{C\alpha_+-A\alpha_-}{BC-AD}.
\end{equation}
The wave function changes under $\phi\rightarrow\phi+2\pi$ as:
\begin{equation}
	\tilde\Psi(\phi+2\pi)=\tilde\Psi(\phi)e^{2\pi im}	.
\end{equation}
Using \eqref{meqDBR} and 
\begin{equation}
	\omega_R=\alpha_++\alpha_-,\qquad \omega_L=\alpha_+-\alpha_-,
\end{equation}
we find that
\begin{equation}
	\tilde\Psi(\phi+2\pi)=\tilde\Psi(\phi)e^{\frac{\pi i}{BC-AD}\left((A+C)\omega_L-(A-C)\omega_R\right)}.
\end{equation}
We can compare this to the relationship
\be\label{temp2DCFT}
\tilde\Psi(\phi + 2\pi) = e^{2\pi i m } \tilde\Psi(\phi) = e^{i4\pi^2 (T_{L} \omega_L-T_{R} \omega_R) } \tilde\Psi(\phi)\,,
\ee
and from this we have the temperatures:
\begin{equation}
	T_L=\frac{A+C}{4\pi(BC-AD)}, \qquad T_R=\frac{A-C}{4\pi(BC-AD)}.
\end{equation}
Thus, we find that the temperatures for the double spinning black ring are:
\begin{equation}
	\begin{split}
		T_L&=	\frac{\sqrt{\Phi } (\mu -\nu ) \left(-\left(1-\mu ^2\right)^{3/2} \nu +\Psi  \sqrt{C_{\mu
					\nu \lambda}  (\nu +1) \Sigma  \Phi  (1-\mu  \nu )}\right)}{4 \pi  \left(\Sigma
			-\left(1-\mu ^2\right) \nu \right) \sqrt{C_{\mu
					\nu \lambda} (\lambda +1)
				\left(1-\mu ^2\right) \nu  \Sigma  \Psi  (1-\mu  \nu )}}\\
		T_R&=	\frac{\sqrt{\Phi } (\mu -\nu ) \left(\left(1-\mu ^2\right)^{3/2} \nu +\Psi  \sqrt{C_{\mu
					\nu \lambda}  (\nu +1) \Sigma  \Phi  (1-\mu  \nu )}\right)}{4 \pi  \left(\Sigma
			-\left(1-\mu ^2\right) \nu \right) \sqrt{C_{\mu
					\nu \lambda} (\lambda +1)
				\left(1-\mu ^2\right) \nu  \Sigma  \Psi  (1-\mu  \nu )}}.\\
	\end{split}
\end{equation}
In the $R\rightarrow\infty$ these reduce to
\begin{equation}
	T_L=\frac{r_++r_-}{4\pi a}, \qquad T_R=\frac{r_+-r_-}{4\pi a},
\end{equation}
as expected.

Now, we can check that we can also map these quantities via the Cardy formula to the Bekenstein-Hawking entropy for the balanced doubly spinning black rings:
\bea
S_+\equiv\frac{\pi^2}{3}(c_{\phi} T_L+c_{\phi} T_R)=\frac{16 \sqrt{2} \pi ^2 R^3 \mu  (1+\nu ) (\mu +\nu )}{(1-\mu ) (1-\mu  \nu )^2}\,,
\eea
and the corresponding value of the inner horizon area is $S_-\equiv\frac{\pi^2}{3}(c_{\phi} T_L-c_{\phi} T_R)$
\bea
S_-\equiv\frac{\pi^2}{3}(c_{\phi} T_L-c_{\phi} T_R)=\frac{16 \sqrt{2} \pi ^2 R^3 \nu  (1+\nu ) (\mu +\nu )}{(1-\nu ) (1-\mu  \nu )^2}.
\eea
 Then the natural choice for the central charge of the black ring (in analogy to Kerr) is 
 \bea
 c_{\phi}=12 J_{\phi}\,.
 \eea
 with angular momentum
 \bea
 J_{\phi}= \frac{8 \pi R^3(\mu+\nu)(1-\mu)}{(1-\mu\nu)^{3/2}}\sqrt{\frac{\nu\lambda(1+\lambda) \Sigma}{(1-\lambda) \Phi \Psi}}\,.
 \eea
This is an appealing, simple picture. In this case, the angular momentum $J_{\phi}$ is present as in the Kerr black hole. Our results are consistent with the CFT identifications for the extremal doubly rotating black ring \cite{Chen:2012yd}. 

Recall from \cite{Chen:2011jb} that the balance condition is
\be
\lambda=\frac{2\,\mu}{1+\mu^2}\,,
\ee
and to obtain exactly the form for the entropy as in \cite{pomeransky2006black} new parameters have to be defined: $\tilde{\lambda}=\mu+\nu$, $\tilde{\nu}=\mu \nu$ and $\tilde{k}=R/\sqrt{2}$. 

Finally, we observe that the area product is quantized,
\be
S_+S_-= 4 \pi^2 J_{\phi }^2 \,,
\ee

One feature of this system is a lack of symmetry between the angular directions; one simply cannot take eigenvalue in the wave function $m=0$ to analyze other sector. Indeed, when the rotation of the $S^2$ is not present it is unclear how to justify the central charge value of $ c_{\phi}$ that is needed to reproduce the entropy of black rings.


\subsection{Boosted Kerr black string}\label{sec:String}

In this subsection we will focus on a family of exact solutions to vacuum 5-dimensional General Relativity that is translationally symmetric. This type of solution is called a Kerr black string. It generalizes a black hole solution but it also extends along a linear $z$-direction in which it can be boosted. The physical parameters include mass, spin and linear momentum $(M, J, P_z)$. The line element for the boosted Kerr black string is given by \cite{Chen:2011jb,pomeransky2006black,Rosa:2012uz} 
\bea
ds^2 &=& - \left(1 - \frac{2 M r \cosh ^2 \sigma}{\Sigma}\right) dt^2  + \left( 1 + \frac{2 M r \sinh ^2 \sigma}{\Sigma} \right) dz^2 \\
&&+  \frac{(\Delta + 2 M r)^2 - \Delta a^2 \sin ^2 \theta}{\Sigma} \sin ^2 \theta d\phi^2 + \Sigma \left( \frac{dr^2}{\Delta} + d \theta^2 \right) \nonumber \\ 
&&+ \frac{4M r}{\Sigma}\left[   \frac{1}{2}\sinh  2\sigma dt dz-  a \sin ^2 \theta \left( \cosh \sigma  dt   -   \sinh \sigma   dz\right) d\phi \right] ,\nonumber 
\eea
where $\Delta = r^2 + a^2 - 2 M r$, $\Sigma = r^2 + a^2 \cos ^2 \theta $. Here $M$ and $a$ can be regarded as the mass and spin and $\sigma$ is the boost parameter. As in Kerr also $M = (r_+ + r_-)/2$, $a = \sqrt{r_+ r_-}$ or $J = M a$. This geometry can be obtained from the unbalanced doubly-spinning black ring solution taking the large $R \rightarrow \infty$ limit. 

The KG equation for the boosted Kerr black string solution is separable under the ansatz 
\be
\tilde\Psi(t,r,\theta,\phi,z)=e^{-i t \omega+i m \phi+i n z} \,\,\tilde\Phi(r,\theta) \equiv e^{-i t \omega+i m \phi+i n z} \,\,\Phi(\theta) \Psi(r).
\ee
The radial wave equation was found in \cite{Dias:2006zv} \:  
\bea\label{DoubleKG}
 \Delta \partial_r (\Delta \partial_r \Psi)- \Delta \left( n^2 r^2 + a^2 \omega^2 - 2 \omega m c_\sigma + \lambda_{lm} \right) \Psi \qquad \qquad \qquad \qquad \qquad \\
 + (2 M r c_\sigma )^2  \Bigg[  \left( \frac{\omega (\Delta + 2 M r)}{2 M r c_\sigma} - \frac{m a}{2 M r }  \right)^2  - \frac{m^2 a^2 \tanh^2_\sigma}{(2 M r)^2}\qquad \qquad \qquad \qquad \nonumber \\
 \qquad \qquad  + \frac{\Delta + 2 M r}{2 M r} \left(   (\omega - n  \, \tanh \sigma)^2  -   \frac{\omega^2}{c_\sigma^2} + \frac{2 \, n\, m \,a \, \tanh \sigma}{c_\sigma(\Delta + 2 M r)} \right)   \Bigg] \Psi = 0\nonumber. 
\eea
We are again interested in the near zone limit of \eqref{DoubleKG} (that is, we focus only on the pole terms). The result is 
\bea
&& \Delta \partial_r(\Delta \partial_r \Psi)  +  (2 M r \cosh \sigma)^2 \Bigg[ \omega -n \tanh \sigma - \frac{m a}{ 2 M r \cosh \sigma} \Bigg]^2 \Psi = 0. \qquad \qquad \qquad
\eea
We can extract the monodromy data analyzing the poles of the differential equation
\bea
\Psi_{\pm} \sim (r-r_\pm)^{ \pm i \alpha_{\pm}}  \qquad \Rightarrow \qquad \alpha_{\pm}=  \frac{2M r_{\pm} \cosh \sigma }{(r_+ -r_-)}  \Bigg[     \omega - n \tanh \sigma  \pm \frac{m a}{2 M r_{\pm}\cosh \sigma} \Bigg]   \nonumber 
\eea
and we can rewrite the KG equation as
\bea
\partial_r(\Delta \partial_r \Psi)  + (r_+ - r_-) \left[ \frac{\alpha^2_+}{r- r_+} - \frac{\alpha^2_-}{r- r_-} \right] \Psi = 0.
\eea
 Having computed the monodromies, considering $n=0$,\footnote{It is worth noting that when $\sigma =0$ the eigenvalue $n$ does not affect the thermodynamics.  This can be compared to the case of the full doubly-spinning black ring, in which the condition $\lambda \to \mu$ results in the Figueras black ring, in which the momentum $J_\psi=0$ \cite{Chen:2011jb}.
 Note this reasoning may also be applied to Section \ref{sec:Dipole}, except there it is the $m$ eigenvalue that decouples in the near zone analysis.
  See the end of Section 4 in \cite{Chen:2011jb}, as well as Section 4 of \cite{Myers:1986un}, for further discussion.
} we are able to the define the frequencies 
\bea
 \omega_L = 2 M  c_\sigma \, \omega \, , \qquad \omega_R = \frac{2 M}{r_+ - r_-} \left(2 M  c_\sigma \, \omega - \frac{m a}{M}\right)\,.
 \eea
 
Likewise, fixing the periodicities along the compact $\phi$-direction
\bea
\Psi(\phi + 2 \pi) = e^{2 \pi i m} \Psi(\phi)= e^{4 \pi^2 i (T_R w_R - T_L w_L)}\Psi(\phi) 
\eea
yields a prescription for the left and right conformal field theory temperatures 
\bea
T_R = \frac{r_+ - r_-}{4 \pi a} \qquad T_L = \frac{r_+ + r_-}{4 \pi a}  .
\eea
The temperatures obey the Cardy relation, with a particular central charge given as:
\bea    S_\pm = \frac{c_\phi  \pi^2}{3}\left(T_L \pm T_R \right) \qquad \text{where} \qquad c_\phi = 12 a M c_\sigma R = 12 J_{\text{BS}}\,.
\eea
For a black string (BS) of length $R$, entropy $S_{\pm}=2 \pi M R \, r_{\pm} c_{\sigma}$ and and spin $J_{\text{BS}}=a M c_\sigma R $ (see e.g. \cite{Dias:2006zv,Rosa:2012uz}).
While the temperatures are independent of the boost parameter, our results show that the central charge picks up this dependence. And, while the results are directly comparable to the $D=4$ Kerr case \eqref{refTemps} when the boost parameter vanishes $(\sigma=0)$, our results indicate that the solution has a more general CFT interpretation than for Kerr. Finally, it is also worth emphasizing that these identifications in the large $R$ limit  are also in full agreement with the results derived in the previous section which further supports our proposals.

Note the system is quantized:
\bea
S_+ S_- = { 4 \pi^2 J_{\text{BS}}^2}.
\eea

%
%

\subsection{Conformal coordinates}


From our discussion of the dipole black ring, we know right away that we will not be able to find conformal coordinates that reproduce \eqref{KGPSring} unless one of two limits is taken: (i) zooming in on one of the black ring horizons or (ii) the black string limit $R\rightarrow\infty$. Limit (i) proceeds exactly as in the dipole black ring case (specifically the discussion surrounding \eqref{dipoleH}), and so we will not repeat that discussion here. In this section we will focus on the second limit $R\rightarrow\infty$.

Let us consider the KG-equation for the doubly-spinning black ring \eqref{DSRadial}. Before taking the $R \rightarrow \infty$ limit, it is convenient to employ a reparametrization 
\bea
\nu= r_-/R, \qquad \mu= r_+/R, \qquad \lambda=r_+ c_{\sigma}/R,
\eea
where we have introduced the Kerr horizons $r_{\pm}$ and $c_{\sigma}$ that will represent the boost parameter in the black string.
Using these definitions and taking the large $R$ limit, the wave equation \eqref{DSRadial} can be written as
\bea
\left[ \partial_r \left(\Delta\,\partial_r\, \right) + \frac{  \left( c_{\sigma} r _+ (r_++r_-)\, \omega -\sqrt{ r _+ r_-} \, m  \right)^2}{(r-r_+) (r_+ - r_-)}-\frac{  \left( c_{\sigma} r _- (r_++r_-)\, \omega -\sqrt{ r _+ r_-} \, m  \right)^2}{(r-r_-) (r_+ - r_-)}\right] \tilde \Phi =0 \,\nonumber
\eea
with $\Delta=(r-r_+)(r-r_-)$. This is none other than the near-region KG-equation for the Kerr black hole that was reported in \cite{Castro:2010fd}, but modified by the boost parameter $c_\sigma$. Changing coordinates to $x=\frac{2r-(r_+ + r_-)}{2(r_+-r_-)}$, we have
\bea
\left[ \partial_x  \bar{\Delta}\,\partial_x + \frac{  \left( c_{\sigma} r _+ (r_++r_-)\, \omega -\sqrt{ r _+ r_-} \, m  \right)^2}{(x-1/2) (r_+ - r_-)^2}-\frac{  \left( c_{\sigma} r _- (r_++r_-)\, \omega -\sqrt{ r _+ r_-} \, m  \right)^2}{(x+1/2) (r_+ - r_-)^2}\right] \tilde \Phi =0 \,.\nonumber
\eea
%
where $\bar{\Delta}=x^2-\frac{1}{4}$. 
Comparing with the Casimir \eqref{CasimirConfCoord}, we obtain conformal coordinates \eqref{xformconfcood} that reproduce the near zone radial equation, with, for example,
\begin{equation}
	\alpha=\frac{r_+ - r_-}{2\sqrt{r_+r_-}}, \qquad \beta=0,\qquad \gamma=\frac{r_+ + r_-}{2\sqrt{r_+r_-}},\qquad \delta=-\frac{1}{c_\sigma(r_++r_-)}.
\end{equation}
Note that these identifications for the conformal coordinates agree with those of Kerr \cite{Perry:2020ndy} by setting the boost parameter $\sigma=0$ where the (unboosted) Kerr black string geometry becomes $ds^2=ds^2_{Kerr}+dz^2$.
At this stage, there is an ambiguity in fixing the conformal coordinates. However, we can make an argument that in general $\alpha=2\pi T_R$ and $\gamma=2\pi T_L$. 
Assuming a Cardy formula and comparing with the entropy for the Kerr black string (see above) we are able to determine the central charge of the CFT dual
\bea
c= 12 a M c_\sigma R=12 J_{\text{BS}}\,.
\eea


\section{Discussion}\label{Discussion}

We have studied the presence of hidden conformal symmetry in five-dimensional systems in which the KG-equation seems non-separable (in the dipole black ring and the doubly spinning black ring cases) as well as their separable large $R$ black string counterparts (in particular the boosted charged black string and the boosted Kerr black string). In analogy to the conjectured non-extremal Kerr/CFT correspondence of \cite{Castro:2010fd}, we propose that a dual CFT exists in a near zone limit of the black ring and black string solutions, and we use a revised monodromy technique to work out the associated CFT left/right temperatures $(T_L , T_R)$ respectively. The present approach provides a derivation of these temperatures based on monodromy data, without the need for a low energy limit. Furthermore, we develop a set of conformal coordinates for each solution we consider, and show that they are related to $T_L$ and $T_R$ in a natural way. Assuming a Cardy entropy formula and comparing with the Bekenstein-Hawking entropy formula we are able to determine the corresponding central charge. This includes the  dipole black ring or string configurations with $c=6 q^3$ and and doubly-spinning black ring or boosted Kerr black string with $c=12 J$.

The identification of left- and right-moving sectors in terms of the monodromy coefficients in higher dimensions proceeded almost exactly as in the Kerr black hole example. The only subtlety is for black ring and strings that the identifications gives a unique way of realizing the hidden conformal symmetry. This is not linked to 5-dimensional spacetimes but rather to the isometries involved. The presence of two commuting $U(1)$ isometries in the 5D Myers-Perry black hole solution gives two inequivalent ways of realizing the hidden conformal symmetry \cite{Krishnan:2010pv,Castro:2013kea}. Likewise, the central charge associated to the associated to the $\psi$ circle is $c_{\psi}=6 J_{\phi}$ and $c_{\phi}=6 J_{\psi}$ for the $\phi$ circle.

We also have the phenomenological observation that the entropy product is independent of the mass, and we can define for black rings and black strings the central charge $c$ as
\bea
\frac{S_{+}S_{-}}{4\pi^2}=\mathcal{F}(J) \rightarrow c\equiv c_R=c_L= 6 \frac{\partial\mathcal{F}}{\partial J}\,.
\eea 

The analysis of present work was inspired in part by the following question: Are all black holes dual to a 2D CFT? Much evidence exists for a Kerr/CFT correspondence both at \cite{Guica:2008mu} and away from \cite{Castro:2010fd} extremality. Furthermore, through studying scattering amplitudes on black hole backgrounds, there is evidence to suggest that 2D CFT duals might exist for more general black hole spacetimes \cite{Hartman:2009nz,Cvetic:2009jn}. Studying the presence of hidden conformal symmetry in more exotic solutions such as black rings and black strings is an ideal arena to push the boundaries of how general a phenomenon it is to have a proposed Near-Horizon/CFT duality.

The black ring and string solutions are also great systems to study the interplay between hidden conformal symmetry and separability. That is, does the presence of a tower of Killing tensors (responsible for the separability of the KG-equation) play a direct role in the presence of hidden \textit{conformal} symmetry in the near zone radial wave equation? We are able to show that a consistent hidden conformal symmetry analysis is possible in \textit{non-separable} systems, provided that we focus on the outer horizon. Focusing only on the outer horizon is essentially what is done to find globally defined hidden symmetry generators, as in \cite{Charalambous:2021mea}.

We would like to stress that the soft hair interpretation of hidden conformal symmetry, as presented for example in \cite{Haco:2018ske}, is not viable in more general contexts, such as black ring solutions. Rather than taking a frequency dependent limit in the wave equation, such as $\omega M<<1$ and $\omega r<<1$, one instead needs to discard all terms of the wave equation except for the poles. For Kerr these two perspectives coincide, but in more general scenarios they do not.

%

\section*{Acknowledgements} 
This work was supported by the NSF grant PHY-2012036 at Utah State University. The work of ABC is partially supported through a Utah NASA Space Consortium Grant fellowship, as well a USU Howard L. Blood Fellowship. VLM is supported by the Icelandic Research Fund under grants 195970-053 
and 228952-051 and by the University of Iceland Research Fund. LFT acknowledges support from USU PDRF fellowship and USU Howard L. Blood Fellowship. MJR is partially supported through the NSF grant PHY-2012036, RYC-2016-21159, CEX2020-001007-S and PGC2018-095976-B-C21, funded by MCIN/AEI/10.13039/501100011033.


\appendix


\section{Black Ring Metrics and Identities}
\label{Appendix:A}

\subsection{Dipole Black Ring}

The following Dipole Black Ring identities are developed from \cite{Emparan:2004wy}. For simplicity, we have worked with the non-dilaton limit $(N=3)$, where $N$ is the dipole charge winding number with magnetic source $A_{\phi}$. The line element is given as:
\begin{multline}
    ds^2=-\frac{F(y)}{F(x)}\left(\frac{H(x)}{H(y)}\right)
\left(dt+C(\nu,\lambda)\: R\:\frac{1+y}{F(y)}\: d\psi\right)^2
\\+\frac{R^2}{(x-y)^2}\: F(x)\left(H(x)H(y)^2\right)\left[
-\frac{G(y)}{F(y)H(y)^3}d\psi^2-\frac{dy^2}{G(y)}
+\frac{dx^2}{G(x)}+\frac{G(x)}{F(x)H(x)^3}d\varphi^2\right]\
\end{multline}
with functions defined as follows:
\begin{align}
    F(\xi)&=1+\lambda\xi,\qquad G(\xi)=(1-\xi^2)(1+\nu\xi)\\
    H(\xi)&=1-\mu\xi.
\end{align}
The curvature of this system involves the existence of energy-momentum sourced by electromagnetic fields. The potential is given by
\begin{equation}
    A_{\phi}=\sqrt{3}C(\nu,-\mu)R\frac{1+\cos{\theta}}{H_{\theta}}+k_1,
\end{equation}
where $H_\theta$ is given in \eqref{Heq}. The constant $k_1$ is associated to the motion of Dirac strings \cite{Emparan:2006mm}. Given the Faraday tensor $F_{\mu \nu}={\partial}_{\mu}A_{\nu}-{\partial}_{\nu}A_{\mu}$, we have that the only non-zero components are
\begin{equation}
    F_{23}=-F_{32}=-{A'}_{\phi}(\theta).
\end{equation}
With this result in hand, we can determine the stress-energy tensor
\begin{equation}
    T_{\alpha \beta}=F_{\mu \alpha}g^{\alpha \beta}F_{\beta \nu}-\frac{1}{4}g_{\mu \nu}F_{\sigma \alpha}g^{\alpha \beta}F_{\beta \phi}g^{\phi \sigma}
\end{equation}
with non zero components
\begin{align}
    T_{00}&=\frac{F_{r}H^{3}_{\theta}{A'}^{2}_{\phi}}{(r^{2}\sin{\theta})^{2}F^{3}_{\theta}H^{9}_{r}}\\
    T_{11}&=\frac{R^{2}(1+\frac{r\cos{\theta}}{R})^{6}H^{3}_{\theta}{A'}^{2}_{\phi}}{(r^{2}\sin{\theta})^{4}F_{\theta}G_{r}H^{6}_{r}}\\
    T_{22}&=\frac{(1+\frac{r\cos{\theta}}{R})^{4}H^{3}_{\theta}\big[-G_{\theta}(1+\frac{r\cos{\theta}}{R})^{2}+2r^{2}F_{\theta}H_{\theta}H^{2}_{r}{\sin{\theta}}^{2}\big]{A'}^{2}_{\phi}}{(r^{3}\sin{\theta})^{2}F_{\theta}G^{2}_{\theta}H^{6}_{r}{\sin{\theta}}^{2}}\\
    T_{33}&=\frac{(1+\frac{r\cos{\theta}}{R})^{4}G_{\theta}\big[2r^{2}G_{\theta}H^{2}_{r}-H^{2}_{\theta}(1+\frac{r\cos{\theta}}{R})^{2}\big]{A'}^{2}_{\phi}}{(r^{3}\sin{\theta}^{2})^{2}F^{2}_{\theta}H^{2}_{\theta}H^{6}_{r}}\\
    T_{44}&=\frac{(1+\frac{r\cos{\theta}}{R})^{6}H^{3}_{\theta}\big[r^{2}F^{2}_{\theta}G_{r}H^{2}_{r}+R^{2}C(\nu,\lambda)(1+\frac{r\cos{\theta}}{R})^{2}(1-\frac{R}{r})^{2}\big]{A'}^{2}_{\phi}}{(r^{2}\sin{\theta})^{4}F^{3}_{\theta}F_{r}H{9}_{r}}\\
    T_{04}&=T_{40}=\frac{RC(\nu,\lambda)(1-\frac{R}{r})(1+\frac{r\cos{\theta}}{R})^{8}H^{3}_{\theta}{A'}^{2}_{\phi}}{(r^{2}\sin{\theta})^{4}F^{3}_{\theta}H^{9}_{r}}
\end{align}
where
\be
F_{r}\equiv F(-R/r)\,,\qquad F_{\theta}\equiv F(\cos\theta)
\ee
\be\label{Heq}
H_{r}\equiv H(-R/r)\,,\qquad H_{\theta}\equiv H(\cos\theta)
\ee
\be 
G_{r}\equiv G(-R/r)\,,\qquad G_{\theta}\equiv G(\cos{\theta})
\ee
\be
C({\sigma}_{1},{\sigma}_{2})=\sqrt{{\sigma}_{2}({\sigma}_{2}-{\sigma}_{1})\frac{1+{\sigma}_{2}}{1-{\sigma}_{2}}}.
\ee

Now we address the stability condition for the ring, balancing the centripetal force against the magnetic repulsion generated between the monopoles distributed in the ring structure. From a geometric perspective, that balance is reached by the avoidance of conical singularities in the $\phi,\psi$ directions, through the constraint
\begin{equation}
    \Delta \phi=2\pi\frac{(1+\mu)^{N/2}\sqrt{1-\lambda}}{1-\nu}
\end{equation}
and conical singularities at $x=-1$ and $y=-1$ , considering the condition
\begin{equation}
    \frac{1-\lambda}{1+\lambda}\Big(\frac{1+\mu}{1-\mu}\Big)^{N}=\Big(\frac{1-\nu}{1+\nu}\Big)^{2}.
\end{equation}

Now we define the extremality conditions. The event horizon and Cauchy horizon for our ring solution are at $r=r_{0}$ and $r=0$, respectively. Extremality is reached when $\nu=0$, that is $r_{0}=0$. As expected, this has consequences on the thermodynamical description of the Dipole Black Ring. Having both the temperature and horizon area given as
\begin{eqnarray}
    &&T=\frac{1}{4\pi R}\frac{{\nu}^{(N-1)/2}(1+\nu)}{(\mu+\nu)^{N/2}}\sqrt{\frac{1-\lambda}{\lambda(1+\lambda)}}\\
    &&{\cal{A}}_{H}=8{\pi}^{2}R^{3}\frac{(1+\mu)^{N}{\nu}^{(3-N)/2}(\mu+\nu)^{N/2}\sqrt{\lambda(1-{\lambda}^{2})}}{(1-\nu)^{2}(1+\nu)},
\end{eqnarray}
by simple inspection we determine that at extremality, $T=0$. In the non-dilatonic limit, $N=3$, the area remains finite at extremality, ${\cal{A}}_{H}\neq0$, leading to degeneracy of the horizon. Then, as a consequence, even at extremality, the entropy $S$ is non-vanishing.

\subsection{Doubly-spinning Black Ring}\label{Appendix:DSBR}

The doubly-spinning black ring identities are extracted from \cite{Chen:2011jb}. The full metric may be represented as
\bea
ds^2 = - \frac{H[y,x]}{H[x,y]} (dt - \omega_\psi d\psi -\omega_\phi d\phi)^2  + \gamma[x,y]\left( \frac{dx^2}{G[x]} +\frac{dy^2}{G[y]}\right)  \nonumber \\
-\frac{1}{H[y,x]}\left( F[x,y] d\psi^2+2 J[x,y] d\psi d\phi +F[y,x] d\phi^2 \right) \ \
\eea
with functions defined by\footnote{Valid solutions to $G_{\mu \nu}[g(\ast, \cdot)] =0 $ are constrained under $0 \leq \nu \leq \mu \leq \lambda < 1$ and $R>0$. The metric is independent of time $-\infty<t<\infty$, angles $0\le \psi,\phi<2\pi$; further, the C-metric-like coordinates $(x,y)$ take ranges $-1< x<1$ and  $-\infty< y<-1$. }:
\begin{align}
	&\gamma[x,y] = \frac{R^2 (1-\mu)^2 (1-\nu) H[x,y]}{(1- \lambda)(1 - \mu \nu ) \Phi \Psi (x-y)^2 } := \frac{R^2 C_{\mu \nu \lambda} H[x,y]}{(x-y)^2}\\ \vspace{7mm}
	&  G[x] = (1-x^2)(1 + \mu x)(1+ \nu x)\\
	& F[x,y]= \frac{R^2}{\mu \nu (1-\mu \nu) \Phi (x-y)^2}  \left(G[x] f_1[\lambda, \mu, \nu; y] + G[y] f_2[\lambda, \mu, \nu; x] \right)\\
	& H[x,y]= (1 - \lambda)(1 - \nu)\Psi \Phi + \lambda (\mu + \nu) \left( (1- \lambda \mu)^2 - (\lambda -\mu)^2 \nu^2 \right) (1+x) \nonumber \\
	&- (1-x^2)(1-y^2) \nu \left( \frac{ \lambda (\lambda-\mu)^2 \nu (\mu + \nu ) (1+x) -(1-\lambda ) \Psi +(\mu +1 - \Phi) }{1-x^2}\right.\\
	&\left.  + \frac{\lambda (\lambda -\mu)(-1 + \lambda \mu) (\mu+\nu)(1+ y)  + \lambda ( \lambda - \mu) (1-\mu) (1-\lambda \mu) (\mu + \nu) + \Sigma \Psi }{1-y^2}  \right. \nonumber \\ &\left. -\left( \Sigma \Psi + \lambda \mu (\lambda - \mu)  (-1 + \lambda \mu)(\mu + \nu) \right)\phantom{\frac{}{}}\right) \nonumber 
\end{align}
\begin{align}
	& J[x,y] =  A[x,y]\left( 1 + \frac{\Sigma \Psi \mu}{\nu} + \frac{\mu (1 + \Sigma \Psi  )}{2}(x+y)  \right.\\
	&\left. +(1+ \mu x )(1+ \mu y) \left(   \Phi (\Phi - (1-\lambda \mu)(2 + \lambda \mu) ) - (1 - \lambda \mu)(\mu^2 -1)   \right) \phantom{\frac{}{}} \right) \nonumber\\
	&  A[x,y] := \nu  \frac{R^2 (\mu + \nu) \sqrt{\nu(\lambda-\mu)(1-\lambda \mu)}(1-x^2)(1-y^2)}{\mu^2 (1 - \mu \nu) \Phi (x-y)}\\
	&\omega_\phi[x,y] = \frac{R(\mu+\nu)  }{ H[y,x]} \sqrt{\frac{\nu \lambda (1-\lambda^2)\Phi \Psi \Sigma}{1 - \mu \nu}} y(1- x^2) \\
	&\omega_\psi[x,y] = \frac{\omega_\phi [x,y]}{\Psi (1-\lambda) y (1-x^2)} \sqrt{\frac{(\lambda - \mu) (1- \lambda \mu)}{\nu  }}\left(\Phi (1+ \nu y)  \right.\\
	&\left. + (1-\mu) \nu (1-y) (1+ x \lambda)+ y \nu (1 - \Phi + \mu)(1-x^2)\right)\nonumber.
\end{align}
Expressions for the functions $f_1$ and $f_2$ can be found in \cite{Chen:2011jb}. We also define 
\begin{equation}
	\Phi \equiv 1 + \mu \nu - \lambda(\mu + \nu) ,\qquad \Psi \equiv \mu(1+ \nu) - \lambda (\nu + \mu^2), \qquad \Sigma \equiv \mu(1- \nu) + \lambda (\nu - \mu^2) .
\end{equation}

Consider the massless, spinless wave equation\footnote{The KG field selection automatically projects onto the lowest spin-weight state(-tower) because the (uncharged) KG field is spin self-dual. Having a fully extended, analytic scalar field is useful in embedding-measured representations of boundary (orbifolds/)manifolds, such as asympotic infinity or coordinate singularities; in particular, the interplay between both (see \cite{Chanson:2020hly,Haco:2018ske,PipolodeGioia:2022exe} ).  } in the background of a doubly-spinning black ring. In particular, we will use the coordinate singular points (event horizons) in order to put the wave equation into a useful form for extracting the near-horizon quasi-normal modes. The KG operator equation is:

\bea
\delta \left[ \mathcal{L}\left[ [\cdot]_{\text{KG}}^{m=0}\right] \right] = 0 \ \Leftrightarrow \  \partial_\mu \left[ \sqrt{|-g|} g^{\mu \nu}\partial_\nu[\cdot]\right] =0.
\eea
Consider a 5D metric with global symmetries in $ x^{\{ \alpha, \beta \} } \in \{ t, \phi, \psi \}$; then $\partial_\alpha [\sqrt{-|g_T|} g^{\mu \nu}] = 0$ and, letting $\{i ,j \} \in \{ x, y\}$ represent the additional coordinates, the (massless) KG equation is:
\bea
0=  \sqrt{|-g|} g^{\alpha \beta}\partial_\alpha \partial_\beta[\cdot]  +\partial_i \left[ \sqrt{|-g|} g^{i j}\partial_j [\cdot]\right] .
\eea 
Then, with a trial solution of $[\cdot] \to e^{i \lambda_\alpha x^\alpha}[ \cdot ]$ $\Rightarrow \ \partial_\alpha \partial_\beta[\cdot] \to \lambda_\alpha \lambda_\beta [\cdot] $ the above may be written as:
\bea
0= - \sqrt{|-g|} g^{\alpha \beta}\lambda_\alpha \lambda_\beta[\cdot]  +\partial_i \left[ \sqrt{|-g|} g^{i j}\partial_j [\cdot]\right]. 
\eea

Further, consider a solution which is poloidally factorized: $[\cdot] \to \varphi_T[t,\phi,\psi] \varphi_P[x,y]  $ where $\varphi_P[x,y] = h[x,y] \tilde\Phi[x,y]$; then $ \partial_a \varphi_P = \varphi_P (\partial_a [ \ln [h \tilde\Phi]] )$, and the poloidal piece of the KG equation becomes:
\begin{align}
	\partial_a [\sqrt{-g} g^{a b}_P \partial_b \varphi ] \ &= \ \frac{R^2\varphi_T }{1-\nu \mu}\partial_a \left[ \frac{ G[x^a] \varphi_P }{(x-y)^2 } \partial_a [ \ln[ h \tilde\Phi ] ] \right]\\
	&= \frac{h R^2\varphi_T }{(1-\nu \mu )(x-y)^2} \left(\partial_a [ G[x^a] \tilde\Phi_{, a} ]  + 2 G[x^a] \left( \partial_a \text{ln}{[h]} - \frac{\delta^{x}_a - \delta^{y}_a}{x-y} \right) \tilde\Phi_{, a}  \right. \nonumber \\
	&\left. + (\partial_a \ln h) G[x^a] \left( \partial_a \ln{[h_{,a} G[x^a]]} - \frac{2(\delta^{x}_a - \delta^{y}_a)}{x-y} \right) \tilde\Phi \right).\nonumber 	
\end{align}
In the above picture the free derivative terms have the same differential envelope ($\partial_a [ G[x^a] [\cdot] ]$), reminiscent of an isotropic fluid polarization. 

Note that $h \to x-y$ gives $\left(  \partial_a \ln{[\frac{h}{x-y}] }\right) \tilde\Phi_{,a}  \to 0$ and  $h_{, a a} \to 0$. In this case, it can be shown that:
\bea
\partial_a [\sqrt{-g} g^{a b}_P \partial_b \varphi ] =  \frac{ R^2 \varphi_T}{(1-\nu \mu )(x-y)}  \left( \partial_a [ G_a \tilde\Phi_{, a} ] - \frac{G_a}{x-y} \left( \frac{2(\delta^{x}_a + \delta^{x}_y))}{x-y} - \frac{ G_{a,a} }{G_a}(\delta^x_a - \delta^y_a) \right) \tilde\Phi \right). \qquad \qquad 
\eea
Then, defining $\Omega[x,y] = \frac{(x-y)^2(1- \nu \mu)}{R^2}\sqrt{|-g|}  = \frac{R^2 C_{\mu \nu \lambda} H[x,y]}{(x-y)^2} \equiv \sqrt{G[x] G[y]|g_P|} $ induces\footnote{Note: $\frac{\Omega_{xy}}{|-g_T|} = \frac{C_{\mu \nu \lambda} H_{xy} (1- \mu \nu)^2 (x-y)^2}{R^2 G_x G_y} $ } the resultant KG-form:

\bea
&&  0 = \partial_a [ G[x^a] \tilde\Phi_{, a} ] - \Omega[x,y]  g^{\alpha \beta} \lambda_\alpha \lambda_\beta \tilde\Phi \\
&&\qquad \qquad \qquad    - \frac{G[x^a]}{x-y} \left( \frac{2(\delta^{x}_a + \delta^{x}_y)}{x-y} - \left( \frac{-2x^a}{1-(x^a)^2} + \frac{\mu}{1+\mu x^a} + \frac{\nu}{1+\nu x^a}\right)(\delta^x_a - \delta^y_a) \right) \tilde\Phi . \nonumber
\eea

From this analysis, it can be shown that
\bea
&&0=\partial_r \left(\left(1-\frac{r^2}{R^2}\right)(r-\mu\,R)(r-\nu\,R)\,\partial_r\, \tilde\Phi \right)+\frac{1}{\sin\theta}\,\partial_\theta \left((1+\mu\cos\theta)(1+\nu\cos\theta)\sin\theta\,\partial_\theta\, \tilde\Phi \right)\nonumber\\
&& \qquad +\left(\frac{C_{\mu\nu\lambda} (1-\mu\nu)^2 \left(1+\frac{r}{R}\cos\theta\right)^2 (r/R)^2 \,\tilde{K}(r,\theta) }{\left(1-\frac{r^2}{R^2}\right)(r-\mu\,R)(r-\nu\,R) (1+\mu\cos\theta)(1+\nu\cos\theta)\sin^2\theta }+\tilde f_r+\tilde f_{\theta}\right)\tilde\Phi  \ \ \ \ 
\eea
where 
\bea
\tilde{K}(r,\theta)&=&n^2 R^2 F_{r\theta}-m^2 F_{\theta r}+2 \,m \,\omega (F_{\theta r}\,\omega_{\phi}-J_{r\theta}\,\omega_{\psi})+2 \,n\,R \,\omega (F_{r\theta }\,\omega_{\psi}+J_{r\theta}\,\omega_{\phi})\nonumber\\
&&- 2\,n\, m\,R \,J_{r\theta}+\omega^2\left(\frac{H_{\theta r}^2}{H_{r\theta}}\,\tilde{F}^2+F_{r\theta}\,\omega_{\psi}^2-F_{\theta r}\,\omega_{\phi}^2+2\,\omega_{\phi}\,\omega_{\psi}\,J_{r\theta}\right) \\
& \text{and} & \ \ \tilde f_r=\frac{(\mu+\nu)\,r}{R}-\frac{2 \,r^2}{R^2}\qquad \tilde f_{\theta}=-\cos\theta\,(\mu+\nu+2\,\mu\,\nu\cos\theta)\,.
\eea




\bibliography{Monodromy}

\providecommand{\href}[2]{#2}\begingroup\raggedright\begin{thebibliography}{10}

\bibitem{Guica:2008mu}
M.~Guica, T.~Hartman, W.~Song, and A.~Strominger, {\it {The Kerr/CFT
  Correspondence}},  {\em Phys. Rev. D} {\bf 80} (2009) 124008,
  [\href{http://arxiv.org/abs/0809.4266}{{\tt arXiv:0809.4266}}].

\bibitem{Castro:2010fd}
A.~Castro, A.~Maloney, and A.~Strominger, {\it {Hidden Conformal Symmetry of
  the Kerr Black Hole}},  {\em Phys. Rev. D} {\bf 82} (2010) 024008,
  [\href{http://arxiv.org/abs/1004.0996}{{\tt arXiv:1004.0996}}].

\bibitem{Carter:1968rr}
B.~Carter, {\it {Global structure of the Kerr family of gravitational fields}},
   {\em Phys. Rev.} {\bf 174} (1968) 1559--1571.

\bibitem{Frolov:2006dqt}
V.~P. Frolov and D.~Kubiznak, {\it {Hidden Symmetries of Higher Dimensional
  Rotating Black Holes}},  {\em Phys. Rev. Lett.} {\bf 98} (2007) 011101,
  [\href{http://arxiv.org/abs/gr-qc/0605058}{{\tt gr-qc/0605058}}].

\bibitem{Frolov:2017kze}
V.~Frolov, P.~Krtous, and D.~Kubiznak, {\it {Black holes, hidden symmetries,
  and complete integrability}},  {\em Living Rev. Rel.} {\bf 20} (2017), no.~1
  6, [\href{http://arxiv.org/abs/1705.05482}{{\tt arXiv:1705.05482}}].

\bibitem{Kubiznak:2006kt}
D.~Kubiznak and V.~P. Frolov, {\it {Hidden Symmetry of Higher Dimensional
  Kerr-NUT-AdS Spacetimes}},  {\em Class. Quant. Grav.} {\bf 24} (2007), no.~3
  F1--F6, [\href{http://arxiv.org/abs/gr-qc/0610144}{{\tt gr-qc/0610144}}].

\bibitem{Emparan:2006mm}
R.~Emparan and H.~S. Reall, {\it {Black Rings}},  {\em Class. Quant. Grav.}
  {\bf 23} (2006) R169, [\href{http://arxiv.org/abs/hep-th/0608012}{{\tt
  hep-th/0608012}}].

\bibitem{Castro:2013kea}
A.~Castro, J.~M. Lapan, A.~Maloney, and M.~J. Rodriguez, {\it {Black Hole
  Monodromy and Conformal Field Theory}},  {\em Phys. Rev. D} {\bf 88} (2013)
  044003, [\href{http://arxiv.org/abs/1303.0759}{{\tt arXiv:1303.0759}}].

\bibitem{Castro:2013lba}
A.~Castro, J.~M. Lapan, A.~Maloney, and M.~J. Rodriguez, {\it {Black Hole
  Scattering from Monodromy}},  {\em Class. Quant. Grav.} {\bf 30} (2013)
  165005, [\href{http://arxiv.org/abs/1304.3781}{{\tt arXiv:1304.3781}}].

\bibitem{Chanson:2020hly}
A.~B. Chanson, J.~Ciafre, and M.~J. Rodriguez, {\it {Emergent black hole
  thermodynamics from monodromy}},  {\em Phys. Rev. D} {\bf 104} (2021), no.~2
  024055, [\href{http://arxiv.org/abs/2004.14405}{{\tt arXiv:2004.14405}}].

\bibitem{Emparan:2004wy}
R.~Emparan, {\it {Rotating circular strings, and infinite nonuniqueness of
  black rings}},  {\em JHEP} {\bf 03} (2004) 064,
  [\href{http://arxiv.org/abs/hep-th/0402149}{{\tt hep-th/0402149}}].

\bibitem{pomeransky2006black}
A.~Pomeransky and R.~Sen'kov, {\it Black ring with two angular momenta},  {\em
  arXiv preprint hep-th/0612005} (2006).

\bibitem{Chen:2011jb}
Y.~Chen, K.~Hong, and E.~Teo, {\it {Unbalanced Pomeransky-Sen'kov black ring}},
   {\em Phys. Rev. D} {\bf 84} (2011) 084030,
  [\href{http://arxiv.org/abs/1108.1849}{{\tt arXiv:1108.1849}}].

\bibitem{Haco:2018ske}
S.~Haco, S.~W. Hawking, M.~J. Perry, and A.~Strominger, {\it {Black Hole
  Entropy and Soft Hair}},  {\em JHEP} {\bf 12} (2018) 098,
  [\href{http://arxiv.org/abs/1810.01847}{{\tt arXiv:1810.01847}}].

\bibitem{Keeler:2021tqy}
C.~Keeler, V.~Martin, and A.~Priya, {\it {Hidden conformal symmetries from
  Killing towers with an application to large-D/CFT}},  {\em SciPost Phys.}
  {\bf 12} (2022), no.~5 170, [\href{http://arxiv.org/abs/2110.10723}{{\tt
  arXiv:2110.10723}}].

\bibitem{Aggarwal:2019iay}
A.~Aggarwal, A.~Castro, and S.~Detournay, {\it {Warped Symmetries of the Kerr
  Black Hole}},  {\em JHEP} {\bf 01} (2020) 016,
  [\href{http://arxiv.org/abs/1909.03137}{{\tt arXiv:1909.03137}}].

\bibitem{Perry:2020ndy}
M.~Perry and M.~J. Rodriguez, {\it {Central charges for AdS black holes}},
  {\em Class. Quant. Grav.} {\bf 39} (2022), no.~4 045009,
  [\href{http://arxiv.org/abs/2007.03709}{{\tt arXiv:2007.03709}}].

\bibitem{Perry:2022udk}
M.~J. Perry and M.~J. Rodriguez, {\it {CFT duals of Kerr-Taub-NUT and beyond}},
   \href{http://arxiv.org/abs/2205.09146}{{\tt arXiv:2205.09146}}.

\bibitem{Bena:2004de}
I.~Bena and N.~P. Warner, {\it {One ring to rule them all ... and in the
  darkness bind them?}},  {\em Adv. Theor. Math. Phys.} {\bf 9} (2005), no.~5
  667--701, [\href{http://arxiv.org/abs/hep-th/0408106}{{\tt hep-th/0408106}}].

\bibitem{Gauntlett:2002nw}
J.~P. Gauntlett, J.~B. Gutowski, C.~M. Hull, S.~Pakis, and H.~S. Reall, {\it
  {All supersymmetric solutions of minimal supergravity in five- dimensions}},
  {\em Class. Quant. Grav.} {\bf 20} (2003) 4587--4634,
  [\href{http://arxiv.org/abs/hep-th/0209114}{{\tt hep-th/0209114}}].

\bibitem{Emparan:2001wn}
R.~Emparan and H.~S. Reall, {\it {A Rotating black ring solution in
  five-dimensions}},  {\em Phys. Rev. Lett.} {\bf 88} (2002) 101101,
  [\href{http://arxiv.org/abs/hep-th/0110260}{{\tt hep-th/0110260}}].

\bibitem{Hawking:1971vc}
S.~W. Hawking, {\it {Black holes in general relativity}},  {\em Commun. Math.
  Phys.} {\bf 25} (1972) 152--166.

\bibitem{Castro:2012av}
A.~Castro and M.~J. Rodriguez, {\it {Universal properties and the first law of
  black hole inner mechanics}},  {\em Phys. Rev. D} {\bf 86} (2012) 024008,
  [\href{http://arxiv.org/abs/1204.1284}{{\tt arXiv:1204.1284}}].

\bibitem{Cardy:1986ie}
J.~L. Cardy, {\it {Operator Content of Two-Dimensional Conformally Invariant
  Theories}},  {\em Nucl. Phys. B} {\bf 270} (1986) 186--204.

\bibitem{Gregory:1993vy}
R.~Gregory and R.~Laflamme, {\it {Black strings and p-branes are unstable}},
  {\em Phys. Rev. Lett.} {\bf 70} (1993) 2837--2840,
  [\href{http://arxiv.org/abs/hep-th/9301052}{{\tt hep-th/9301052}}].

\bibitem{Wiseman:2002ti}
T.~Wiseman, {\it {From black strings to black holes}},  {\em Class. Quant.
  Grav.} {\bf 20} (2003) 1177--1186,
  [\href{http://arxiv.org/abs/hep-th/0211028}{{\tt hep-th/0211028}}].

\bibitem{Horowitz:2001cz}
G.~T. Horowitz and K.~Maeda, {\it {Fate of the black string instability}},
  {\em Phys. Rev. Lett.} {\bf 87} (2001) 131301,
  [\href{http://arxiv.org/abs/hep-th/0105111}{{\tt hep-th/0105111}}].

\bibitem{Compere:2010fm}
G.~Compere, S.~de~Buyl, S.~Stotyn, and A.~Virmani, {\it {A General Black String
  and its Microscopics}},  {\em JHEP} {\bf 11} (2010) 133,
  [\href{http://arxiv.org/abs/1006.5464}{{\tt arXiv:1006.5464}}].

\bibitem{Elvang:2007hs}
H.~Elvang and M.~J. Rodriguez, {\it {Bicycling Black Rings}},  {\em JHEP} {\bf
  04} (2008) 045, [\href{http://arxiv.org/abs/0712.2425}{{\tt
  arXiv:0712.2425}}].

\bibitem{Krishnan:2010pv}
C.~Krishnan, {\it {Hidden Conformal Symmetries of Five-Dimensional Black
  Holes}},  {\em JHEP} {\bf 07} (2010) 039,
  [\href{http://arxiv.org/abs/1004.3537}{{\tt arXiv:1004.3537}}].

\bibitem{Chen:2012yd}
B.~Chen and J.-j. Zhang, {\it {Holographic Descriptions of Black Rings}},  {\em
  JHEP} {\bf 11} (2012) 022, [\href{http://arxiv.org/abs/1208.4413}{{\tt
  arXiv:1208.4413}}].

\bibitem{Rosa:2012uz}
J.~G. Rosa, {\it {Boosted black string bombs}},  {\em JHEP} {\bf 02} (2013)
  014, [\href{http://arxiv.org/abs/1209.4211}{{\tt arXiv:1209.4211}}].

\bibitem{Dias:2006zv}
O.~J.~C. Dias, {\it {Superradiant instability of large radius doubly spinning
  black rings}},  {\em Phys. Rev. D} {\bf 73} (2006) 124035,
  [\href{http://arxiv.org/abs/hep-th/0602064}{{\tt hep-th/0602064}}].

\bibitem{Myers:1986un}
R.~C. Myers and M.~J. Perry, {\it {Black Holes in Higher Dimensional
  Space-Times}},  {\em Annals Phys.} {\bf 172} (1986) 304.

\bibitem{Hartman:2009nz}
T.~Hartman, W.~Song, and A.~Strominger, {\it {Holographic Derivation of
  Kerr-Newman Scattering Amplitudes for General Charge and Spin}},  {\em JHEP}
  {\bf 03} (2010) 118, [\href{http://arxiv.org/abs/0908.3909}{{\tt
  arXiv:0908.3909}}].

\bibitem{Cvetic:2009jn}
M.~Cvetic and F.~Larsen, {\it {Greybody Factors and Charges in Kerr/CFT}},
  {\em JHEP} {\bf 09} (2009) 088, [\href{http://arxiv.org/abs/0908.1136}{{\tt
  arXiv:0908.1136}}].

\bibitem{Charalambous:2021mea}
P.~Charalambous, S.~Dubovsky, and M.~M. Ivanov, {\it {On the Vanishing of Love
  Numbers for Kerr Black Holes}},  {\em JHEP} {\bf 05} (2021) 038,
  [\href{http://arxiv.org/abs/2102.08917}{{\tt arXiv:2102.08917}}].

\bibitem{PipolodeGioia:2022exe}
L.~Pipolode~Gioia and A.-M. Raclariu, {\it {Eikonal Approximation in Celestial
  CFT}},  \href{http://arxiv.org/abs/2206.10547}{{\tt arXiv:2206.10547}}.

\end{thebibliography}\endgroup
\bibliographystyle{jhep}

\end{document}